\documentclass[10pt,conference]{IEEEtran}
\usepackage{setspace}
\usepackage{subfigure}
\usepackage{cite}
\usepackage{amsmath,amssymb,amsfonts}
\usepackage{algorithmic}
\usepackage{graphicx}
\usepackage{textcomp}
\usepackage{xcolor}
\usepackage{subfigure}
\usepackage{braket}
\usepackage[T1]{fontenc}
\usepackage{xcolor}
\usepackage{float}
\usepackage{physics}
\ifCLASSINFOpdf
\else
\fi
\usepackage{dblfloatfix}

\begin{document}
%
\title{Optimistic Entanglement Purification in Quantum Networks}

\author{

\IEEEauthorblockN{Mohammad Mobayenjarihani\IEEEauthorrefmark{1}, Gayane Vardoyan\IEEEauthorrefmark{2}, Don Towsley\IEEEauthorrefmark{1}}
\IEEEauthorblockA{
\IEEEauthorrefmark{1}Manning College of Information and Computer Sciences, University of Massachusetts, Amherst\\
\IEEEauthorrefmark{2}QuTech and Faculty of Electrical Engineering, Mathematics and Computer Science, Delft University of Technology\\
mobayen@cs.umass.edu, g.s.vardoyan@tudelft.nl, towsley@cs.umass.edu}




}


%


\maketitle
\thispagestyle{plain}
\pagestyle{plain}


\begin{abstract}

Noise and photon loss encountered on quantum channels pose a major challenge for reliable entanglement generation in quantum networks.
In near-term networks, heralding is required to inform endpoints of successfully generated entanglement. If after heralding, entanglement fidelity is too low, entanglement purification can be utilized to probabilistically increase fidelity. Traditionally, purification protocols proceed as follows: generate heralded EPR pairs, execute a series of quantum operations on two or more pairs between two nodes, and classically communicate results to check for success. Purification may require several rounds while qubits are stored in memories, vulnerable to decoherence. In this work, we explore the notion of optimistic purification in a single link setup, wherein classical communication required for heralding and purification is delayed, possibly to the end of the process. Optimism reduces the overall time EPR pairs are stored in memory. While this is beneficial for fidelity, it can result in lower rates due to the continued execution of protocols with sparser heralding and purification outcome updates. We apply optimism to the entanglement pumping scheme, ground- and satellite-based EPR generation sources, and current state-of-the-art purification circuits. We evaluate sensitivity
performance to a number of parameters including link length, EPR source rate and fidelity, and memory coherence time. We observe that our optimistic protocols are able to increase fidelity, while the traditional approach becomes detrimental to it for long distances. 
We study the trade-off between rate and fidelity under entanglement-based QKD, and find that optimistic schemes can yield higher rates compared to non-optimistic counterparts, with most advantages seen in scenarios with low initial fidelity and short coherence times.
\end{abstract}


%

\IEEEpeerreviewmaketitle
\section{Introduction}
\label{sec:intro}
Certain features of quantum mechanics, such as superposition, entanglement, and interference, have the potential to equip us with applications that are not achievable in the classical world. Examples of quantum-enabled advantages include exponential and polynomial algorithmic speedups\cite{book:nielsonanchuang} and provably secure communication~\cite{paper:qkd}.
Besides being able to provide the latter, quantum networks~\cite{paper:quantuminternet} can also support distributed quantum computation~\cite{paper:distributedqcomputing}, clock synchronization~\cite{paper:clocksynch}, and quantum sensing~\cite{paper:quantum_sensing}. An essential requirement for distributed quantum applications is entanglement of sufficiently high quality shared between nodes. Consequently, one of the main goals of a quantum network is to reliably distribute this resource across a potentially large distance.

A maximally entangled bipartite state (also known as an Einstein-Podolsky-Rosen (EPR)~\cite{paper:eprpair} or Bell pair) is a pair of qubits that are entangled such that if we measure the quantum state of one, then we know the exact state of the other. One can use photons to generate and distribute EPR pairs but due to exponential photon loss in optical fiber, the generation of an EPR pair over a long distance poses a significant challenge. Further, due to the No-Cloning Theorem~\cite{paper:noclone}, one cannot copy or amplify quantum information at intermediate stations.
A solution is to use quantum repeaters~\cite{paper:inside_quantum_repeaters,paper:dbcz}  that assist with long-distance entanglement generation via entanglement swapping \cite{paper:teleport_swap, paper:teleport_experiment}.

\begin{figure}
    \centering
    \includegraphics[width=0.95\linewidth,trim={0cm 9cm 0cm 7cm},clip]{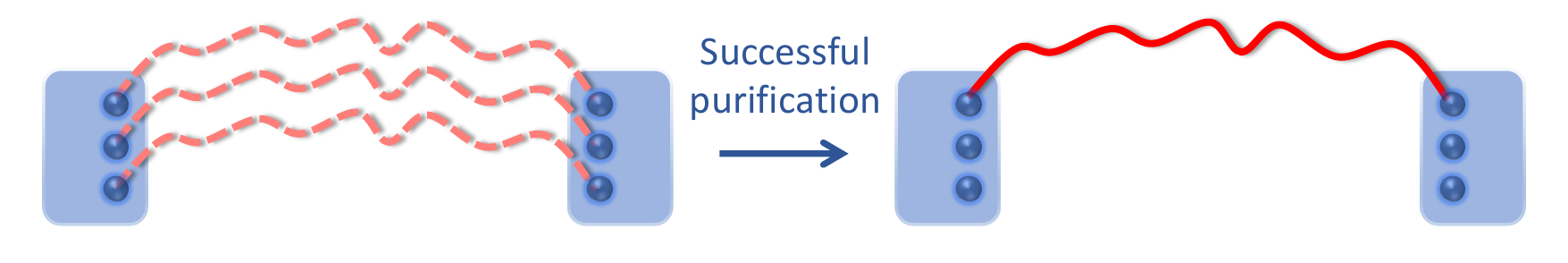}
    \caption{Purification example. Two nodes, with three quantum memories each, begin with three imperfect entangled states (red curves, dashed). After purification is carried out (successfully in the example), the nodes are left with a higher-quality single entangled state (red curve, solid).}
    \label{fig:purif_high_level}
\end{figure}
Imperfect memories, decoherence, and gate noise preclude the distribution of perfect entanglement within a quantum network. In reality, what nodes receive are low quality EPR pairs.
Entanglement quality is crucial for distributed quantum applications, \textit{e.g.}, quantum key distribution (QKD)\cite{paper:qkd,paper:qkd_e91}, Blind Quantum Computation (BQC)~\cite{paper:blind_computation}, as it can determine not only performance measures specific to such an application, but also the feasibility of carrying it out at all.
It is therefore necessary to take heed of and increase this quality when possible. One measure of entanglement quality is \textit{fidelity}, which quantifies the closeness of a given quantum state to some desired state. In quantum networks, a commonly sought-after goal is the distribution of high-fidelity entanglement, where fidelity is computed in reference to one of the four Bell pairs. One way to increase fidelity is through purification~\cite{paper:purification_first},
which involves the application of local gates and measurements on both ends of a shared entangled state, followed by classical information exchange to communicate success or failure of this probabilistic process. Figure~\ref{fig:purif_high_level} illustrates the method at a high level.
\begin{figure}[t]
    \centering
    \includegraphics[scale=0.35]{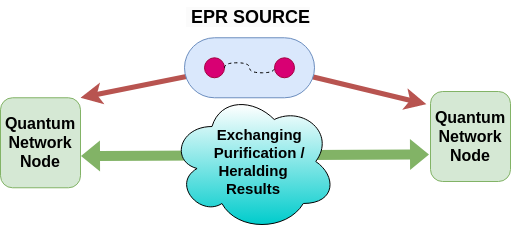}
    \vspace{-10pt}
    \caption{The setup is comprised of an entanglement source situated between two quantum network nodes. The nodes, capable of performing purification, are equipped with quantum memories that can store entangled states.}
    \label{fig:general_scheme}
\end{figure} 

Heralded entanglement purification (HEP) is a necessary mechanism for first-generation quantum networks~\cite{paper:optimal_architecture }, and yet, practical execution workflows for such protocols still require more study. Our work investigates the advantages and limitations of purification for two nodes connected by a single quantum link -- a building block for quantum networks -- as a means of improving our understanding of how such workflows could be designed and realized on a fully fledged network. Figure~\ref{fig:general_scheme} illustrates the setting we consider: two nodes are connected via a classical channel used for heralding entanglement and exchanging purification results. 
Equidistant from both nodes is an entanglement generation source that distributes sub-unit fidelity entanglement.
Nodes are equipped with imperfect quantum memories and noisy quantum gates.

Purification can be performed by two nodes that share at least two heralded entangled pairs: 
one, which we denote as the \textit{main} pair, is kept,
while others, often called \textit{sacrificial} pairs, are eventually measured\footnote{More generally, $n\to k$ purification, with $n>k$ initial and $k$ resulting states, is also possible.} \cite{paper:purification_first, paper:duetch, paper:dbcz, paper:pumping}. The traditional way of carrying out purification involves each node performing local operations on its qubits and measuring all sacrificial pairs. Then, based on measurement results, which are exchanged over a classical channel, purification is deemed either successful or unsuccessful. Upon success, the parties may perform further purification on the purified entangled state(s), or allow an application to consume the entanglement. In case of failure, the nodes are forced to discard the main pair and begin the entire process anew. In this paper, we refer to the traditional method as the \textit{baseline protocol (BASE)}.

Classical communication -- a required part of purification, is a potentially significant cause of fidelity degradation -- 
the main entangled pairs must remain in noisy storage while awaiting confirmation. If a purification scheme has several rounds  (\textit{e.g.}, the pumping scheme~\cite{paper:dbcz}), or each purification circuit includes several measurements, then all results must be checked~\cite{paper:purification_last_stefan}, and this further increases the storage time of a pair. Checking purification results costs time at least equal to the data propagation delay on a link. This makes traditional purification impractical for longer distances when the nodes involved are equipped with noisy quantum memories.  

A characteristic property of the purification schemes that we study is the reduced wait time of stored entanglement via curtailment of overall classical communication.  This reduced storage time in turn impacts the fidelity of entangled pairs by the time they are ready to be consumed by an application. 
An example of such a scheme is one that foregoes a number of classical communication rounds, continuing on to further purification steps without checking for purification success/failure.
This idea was introduced by Hartmann \emph{et al.} in~\cite{paper:blind_purification}. The authors applied their idea to heralded EPR pairs in the pumping scheme \cite{paper:blind_purificaiton_2nd}, and showed that nodes can be optimistic with respect to purification results, checking purification outcomes only at the end. In this work, we refer to their scheme as the \textit{heralded-optimistic protocol (HOPT)} .

In this work, we further increase optimism, by applying it not only to purification results but also to heralding signals. Intuitively, our optimistic protocol (OPT) can yield even higher fidelities since entanglement spends even less time in quantum memory. Similar to the work in \cite{paper:blind_purificaiton_2nd}, we apply our optimistic approach to the pumping scheme, in ground- and satellite-based setups\cite{paper:satellite_china}, and show that for large distances and short memory coherence times, our approach increases fidelity while HOPT and BASE can harm fidelity.
Nevertheless, a heightened degree of optimism can decrease the overall rate, 
since more entanglement will be spent on failed purification procedures.
Thus, a trade-off exists between rate and fidelity. We study this rate-fidelity trade-off with the secret key rate (SKR) of the BB84  protocol \cite{paper:qkd_e91, paper:qkd}. We evaluate the SKR on the pumping scheme~\cite{paper:dbcz},  for a range of hardware parameters including the link's entanglement generation rate, the initial fidelity of generated entanglement, and quantum memory coherence time. We also study the effect of distance between nodes on the secret key rate. We also evaluate a current state-of-the-art purification circuit~\cite{paper:purification_last_stefan} that includes multiple purification checkpoints. We observe that in harsh environments -- lower initial (pre-purification) fidelity and short coherence time -- optimistic schemes are advantageous for the SKR.
In scenarios with higher coherence times, one may switch to the baseline or the heralded-optimistic protocol, and in the case of high initial fidelity, purification may not be necessary at all.

The remainder of this paper is structured as follows: in Section~\ref{section:related_work} we discuss related work in purification schemes. In Section~\ref{section:background} we provide the necessary background in quantum networking. In Section~\ref{section:methodology}, we explain our optimistic approach and methodology. In Section~\ref{section:evaluation}, we evaluate our optimistic approach on a number of different purification schemes. Finally, in Section~\ref{section:conclusion}, we conclude our work and discuss challenges and future directions.

\section{Related Work}
\label{section:related_work}
Entanglement purification was introduced by Bennett~\emph{et al.} in~\cite{paper:purification_first}.
They developed a circuit to improve the fidelity of one Werner state \cite{paper:werner_state} (see Section~\ref{section:background} for a definition of this state) by sacrificing another state of the same form.
In this work, the authors did not evaluate purification performance in terms of rate and fidelity in quantum networks. Further, the effects of memory coherence and entanglement storage time on state fidelity and secret key rate were not considered. 
Deutsch \emph{et al.}~\cite{paper:duetch} improved previous work by proposing a protocol -- often referred to as DEJMPS -- which converges faster and requires fewer resources. The scheme does not restrict the initial states to be Werner -- in this relaxation, a state can be any linear combination of  Bell basis states. In \cite{paper:duetch}, there is no evaluation of the effect of memory noise on final fidelity. D{\"u}r \emph{et al.}~\cite{paper:dbcz} proposed the pumping scheme (see Section~\ref{section:methodology}) and the application of purification in quantum repeaters~\cite{paper:bdcz_wrong_ciatation,paper:dbcz}; however, they did not consider quantum memory storage noise in their analysis.

Hartmann \emph{et al.} studied the effect of memory noise on quantum repeaters with purification in \cite{paper:blind_purification}. Their noise model accounts for noisy two-qubit gates and dephasing in quantum memories. They proposed nodes perform DEJMPS purification and entanglement swapping without checking results or applying corrections until the very end in a quantum network -- a manner of operation they dubbed \textit{blind mode}. In this work, we show that this methodology can exacerbate state fidelity when distances are large and memory coherence times are short. In \cite{paper:blind_purificaiton_2nd}, the authors analyzed the scalability of blind repeaters, while still \emph{heralding} EPR pair generation. In our work, we show that we can also be optimistic about \emph{heralding} signals, thereby improving performance in terms of fidelity and SKR.

All the aforementioned papers apply sub-optimal purification circuits. Nickerson \emph{et. al} introduced the STRINGENT protocol, which outperforms previous protocols in terms of fidelity improvement and quantum state consumption \cite{paper:optimized_stringent}. The effects of waiting times (arising from delays due to classical communication of heralding and purification results)
on quantum states were not evaluated, however. 
Krastanov \emph{et. al} applied a genetic algorithm to optimize purification circuits with respect to resource consumption and output fidelity \cite{paper:purification_last_stefan}. The algorithm takes the initial state fidelity and the maximum allowed number of operations as input parameters. In their work, circuit performance evaluation did not consider the effect of classical communication-induced waiting time and storage noise on output fidelity and rate. As the results in \cite{paper:purification_last_stefan} are the current state-of-the-art in purification, we apply our optimistic scheme to these circuits and evaluate output fidelity and overall rate, while also incorporating storage noise and classical communication time overhead.

\section{Quantum Networking Background}
\label{section:background}
In this section, we provide necessary quantum background for this paper. We begin by introducing EPR pairs, fidelity, quantum channels, the secret key fraction of BB84, quantum repeaters, and purification in more detail. We also explain the entanglement generation setup that we use throughout this work.
 
\subsection{EPR Pairs}
 EPR pairs (also known as Bell states~\cite{book:quantum_nielson}) are the following two-qubit quantum states: $\ket{\phi^{\pm}} = (\ket{00} \pm \ket{11})/\sqrt{2} $ and $\ket{\psi^{\pm}} = (\ket{01} \pm \ket{10})/\sqrt{2}$.  A common objective for nodes in a quantum network is to be in possession of one qubit of a Bell state, \textit{e.g.}, $\ket{\phi^+} = (\ket{00} + \ket{11})/\sqrt{2}$, with the other qubit belonging to another node with whom an application is jointly being carried out. 
 
 \subsection{Fidelity and Noise Model}
 Fidelity is a quantity that measures the closeness of two quantum states. Given a density matrix $\rho$ of a non-maximally entangled bipartite state, the fidelity $F\in[0,1]$ with reference to $\ket{\phi^+}$ is given by\footnote{We note that another widely accepted definition of fidelity employs a square root.}
\begin{align}
    F(\rho) = \bra{\phi^+}\rho\ket{\phi^+},
\end{align}

clearly, higher values are desirable.\\
In the real world, quantum gates and quantum memories are imperfect and may inadvertently apply noise to qubits, decreasing their fidelities. In this work, we consider the effect of noisy two-qubit gates on qubits, where noise is modeled by a depolarization channel for quantum gates. Namely, upon application of a two-qubit quantum gate $U$ on the density matrix $\rho$ of an $n$-qubit system, the transformation is successful with probability $p_g$, and the two qubits undergoing the transformation are depolarized with probability $1-p_g$:

\begin{align}
\label{eq:2qbit_depolarization}
    \rho' = p_gU\rho U^{\dagger} + (1-p_g)Tr_{i,j}(\rho)\otimes\dfrac{I}{4},
\end{align}
where $\rho'$ is the resulting density matrix, $Tr_{i,j}$ is a partial trace over qubits $i$ and $j$ that are affected by $U$, and $I$ is the identity matrix. In this work, we assume that controlled gates (\textit{e.g.}, CNOT and CZ) depolarize both control and target qubits, while single-qubit gates are assumed to be ideal.

Similar to quantum gates, measuring a qubit introduces errors in the output state. Measurement can project an arbitrary state to the correct state with probability $p_m$ or to the wrong state with probability $1-p_m$. An example is an imperfect projection onto the $\ket{0}$ state: 
\begin{align} \label{eq:imperfect_projection}
    \rho' = p_m\ketbra{0}{0}\rho\ketbra{0}{0} + (1-p_m)\ketbra{1}{1}\rho\ketbra{1}{1},
\end{align}
where $\rho$ and $\rho'$ 
are the density matrices of the pre- and post-measurement states, respectively.

We also account for the time-dependent noise affecting qubits stored in quantum memories.
We assume that this noise is described by two types of errors: amplitude damping and dephasing. Amplitude damping is associated with the parameter $T_1$, which characterizes how rapidly a state loses its excitation, and dephasing is associated with the parameter $T_2$ which describes how rapidly a state loses its phase information~\cite{paper:linklayer, paper:netsquid}. The amplitude damping channel acts as follows on the density matrix $\rho$:
\begin{align}
\label{eq:amplitudedamping}
     &\rho \mapsto E_0\rho E_0^{\dagger} + E_1\rho E_1^{\dagger},\\
    &E_0 = \ketbra{0}{0} + \sqrt{1-\lambda}\ketbra{1}{1}, \nonumber\\
    &E_1= \sqrt{\lambda}\ketbra{0}{1}, \nonumber
\end{align}
where $\lambda = 1-e^{-t/T_1}$ and $t$ is the time that the qubit is stored in memory. The stored qubit then goes through a dephasing channel that acts as follows:
\begin{align}
\label{eq:dephasing}
     &\rho \mapsto (1-p_z)\rho + p_zZ\rho Z\\
     &p_z = \dfrac{1}{2}\left(1-e^{-t/T_2}e^{t/(2T_1)}\right), \nonumber
\end{align}
where $Z$ is the Pauli $Z$ gate and $t$ is the time that the qubit spends in the memory. The composition of amplitude and phase damping as described above is thought to be a generally effective way to model state evolution in quantum memories (see discussion in \cite{coopmans2021tools} and references therein).

Another error that can occur is photon loss, one of the main obstacles in a quantum network. The probability of successfully transmitting a photon over optical fiber depends on the fiber transmissivity $\eta_f$. The latter decreases exponentially with distance (or link length) $l$. The probability of transmitting a photon over distance $l$ is
\begin{align}
\label{eq:photon_loss}
\eta_f = 10^{(-\alpha_f\times l)/10},
\end{align}
where $\alpha_f$ is the fiber attenuation coefficient \cite{paper:netsquid}.

\subsection{Secret Key Fraction}
A direct application of EPR pairs is entanglement-based QKD such as entanglement-based BB84 and the E91 protocol~\cite{paper:qkd_e91}. 
The secret key rate of BB84 is an increasing function of entanglement rate and fidelity. 
Recall that purification sacrifices EPR pairs to increase a target state's fidelity. This  has the effect of reducing the entanglement generation rate, thus manifesting a rate-fidelity trade-off problem that makes it difficult to decide whether purification is beneficial. Fidelity influences the secret key rate via the
secret key fraction, $SKF_{BB84}$, given by 
\begin{align}
\label{eq:skf_bb84}
    SKF_{BB84} &= \max(1-h(\theta_x)-h(\theta_z),0)\\
    ~\text{where  } \nonumber \theta_x &= Tr(\rho X \otimes X),\quad \theta_z = Tr(\rho Z \otimes Z),
\end{align}
 $X$, $Z$ are the Pauli $X$ and $Z$ operators, respectively; $Tr$ is the matrix trace, and $h(p) = -p\log(p)-(1-p)\log(1-p)$ is the binary entropy~\cite{paper:qkd_secret_key_fraction}. The secret key rate (SKR) is the production of $SKF_{BB84}$ and the rate of EPR pairs with the density matrix of $\rho$.  We later study the rate-fidelity trade-off of different purification schemes via secret key rate.

\subsection{EPR Pair Generation Model and Purification}
\begin{figure}
    \centering
    \includegraphics[scale=0.5]{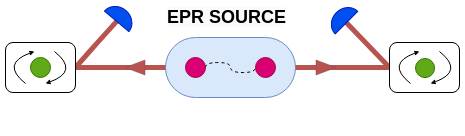}
    \vspace{-5mm}
    \caption{EPR pair generation setup. The source in the middle sends half of an EPR pair to each quantum network node. Each node entangles its photon with an atom in a cavity and measures the photon, or, in case of failure, heralds photon loss to the other node. }
    \vspace{-10pt}
    \label{fig:midsource}
\end{figure}

Figure~\ref{fig:midsource} illustrates the entanglement generation setup considered in this work: a source located between two network nodes distributes entanglement, with polarization encoding used on the photons of each state~\cite{paper:inside_quantum_repeaters}.  An implementation of this abstracted EPR pair generation scheme is introduced in~\cite{paper:spdc_prajit}.
In this scheme, each node has an atom in a cavity. We label photons $p_1$ and $p_2$ and atoms $a_1$ and $a_2$, where each subscript represents the node to which these resources belong. The source distributes states of the form
\begin{align*}
\ket{\phi^+}_{p_1p_2} = (\ket{00}_{p_1p_2} +\ket{11}_{p_1p_2})/\sqrt{2},
\end{align*}
where horizontal polarization for $p_i$ is represented by $\ket{0}_{p_i}$ and vertical polarization by $\ket{1}_{p_i}$.  Here, we assume each attempt to generate a $\ket{\phi^+}_{p_1p_2}$ state is successful at the source.
Each photon $p_i$ is then transmitted to atom $a_i$ located at node $i$. Each atom begins in a superposition of the ground and excited states:
\begin{align}
\ket{+}_{a_i} = 
(\ket{0}_{a_i} + \ket{1}_{a_i})/\sqrt{2},
\end{align}
where $\ket{0}_{a_i}$ represents the ground state and $\ket{1}_{a_i}$ the excited state. After receiving a photon, each node applies a $CZ$ operation on the photon and the atom, bringing the overall state to 

\begin{align}
    \ket{\psi} = &1/2\ket{\phi^+}_{a_1a_2}\otimes[\ket{00}_{p_1p_2} + \ket{11}_{p_1p_2}] +\\  &1/2\ket{\psi^+}_{a_1a_2}\otimes[\ket{00}_{p_1p_2} - \ket{11}_{p_1p_2}]. \nonumber
\end{align}
Both nodes then measure their photons in the diagonal basis (\textit{i.e}, $\{\ket{+},\ket{-}\}$ basis), and apply corrections on the resulting EPR pair based on the measurement results.

In this last stage, upon photon measurement, a node applies the Pauli $X$ gate on its atomic qubit if and only if it observed $\ket{+}$ as the outcome. 
Once an EPR pair is established, it may be consumed directly by an application, \textit{i.e.}, without any purification; we say in this case that the nodes have performed \textit{direct sharing} of entanglement.

In the introduction, we described purification at a high level; here we elaborate more. As previously mentioned, the purpose of purification is to increase the fidelity of a shared entangled pair between two nodes in a quantum network.
Bennett~\emph{et al.} introduced the first purification scheme in~\cite{paper:purification_first}, sometimes called the BBPSSW protocol.
In this proposal, one Werner state, \textit{i.e.}, a state that can be expressed as

\begin{align}
\label{eq:werner}
    \rho = \dfrac{4F_0-1}{3}\ketbra{\phi^+}{\phi^+} + \dfrac{1-F_0}{3}I_4,
\end{align}
with $F_0$ its initial fidelity~\cite{paper:werner_state},
is sacrificed to increase the fidelity of another. In this work, we assume our entanglement generation mechanism generates Werner states as in see \eqref{eq:werner}.
Since noisy gates and noisy quantum state storage may result in a mixed state that is not Werner, it is often more accurate to relax the Werner assumption and allow input states to be a linear combination of Bell states.

\begin{figure}
    \centering
    \includegraphics[scale = 0.15]{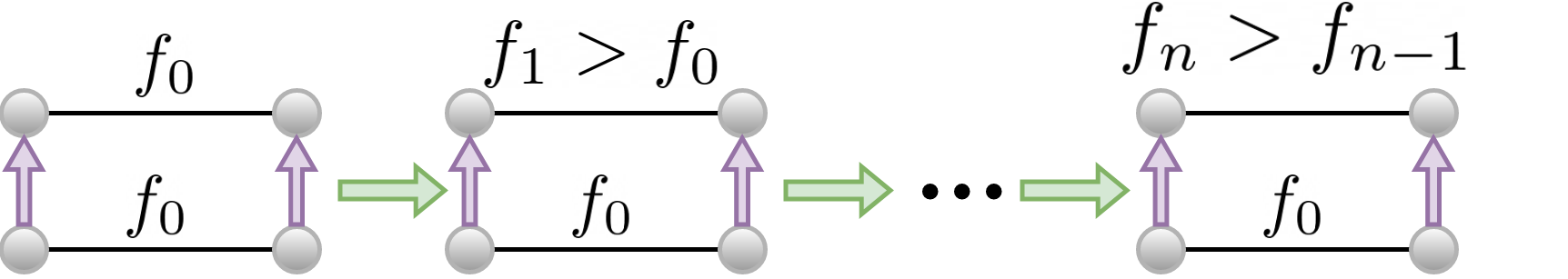}
    \caption{An illustration of the entanglement pumping purification scheme with $n+1$ purification steps. The nodes begin with two EPR pairs with equal fidelity $f_0$, and pump the main EPR pair (top) with sacrificial EPR pairs (bottom) until purification stops yielding significant benefits.}
    \label{fig:pumping}
    \vspace{-10pt}
\end{figure}

For such states, the DEJMPS protocol introduced in ~\cite{paper:duetch} outperforms BBPSSW.
DEJMPS can be applied successively to the same EPR pair to further increase its fidelity. Such a procedure can be carried out by the pumping scheme introduced by D{\"u}r \emph{et al.} in \cite{paper:dbcz}. The method increases a main EPR pair's fidelity by consecutively purifying it with another sacrificial EPR pair. In this work, we refer to each purification of the main EPR pair by a sacrificial EPR pair as a \textit{purification step}; Figure~\ref{fig:pumping} illustrates this process that includes $n+1$ purification steps. 
Note that with the pumping scheme, the fidelity of the main EPR pair ceases to improve after a number of steps that depend on the sacrificial pair fidelity.
Finally, as before, nodes exchange purification results via a classical channel to determine if purification was successful or not. If any purification round fails, the entire process must be restarted.

Although the aforementioned purification techniques increase entanglement fidelity, they are not optimized to use as few EPR pairs as possible or to yield the highest fidelity improvement. Optimization techniques can be applied to purification schemes to 
address these shortcomings.
Krastanov \emph{et al.} applied a genetic algorithm to optimize purification circuits in \cite{paper:purification_last_stefan}. For the noise model in their generated circuits, they considered imperfect measurement projection as in~\eqref{eq:imperfect_projection} and depolarization in controlled gates as in~\eqref{eq:2qbit_depolarization}. Some of these circuits include several projective measurements and require
the nodes to classically communicate results as part of the protocol. In our work, we evaluate our optimistic protocol on the traditional pumping scheme, as well as on an optimized circuit from \cite{paper:purification_last_stefan}.
\subsection{Satellite Setup}
While optical fiber transmissivity decreases exponentially with link length, in free space, this decrease follows a polynomial trend. Consequently, the use of satellites~\cite{paper:satellite_china,paper:spooky} and photon transmission through free space have gained significant attention as emerging technologies that appear to make EPR pair distribution over long distances more feasible.
Nevertheless, due to longer propagation delays for classical messages, EPR pair distribution with satellite technology potentially
introduces longer waiting times for stored quantum states. Stored entangled pairs thus suffer more decoherence, 
suggesting that such an entanglement generation setting could benefit from a reduction of overall classical communication.

\begin{figure}[t]
\centering
    \includegraphics[width=0.95\linewidth]{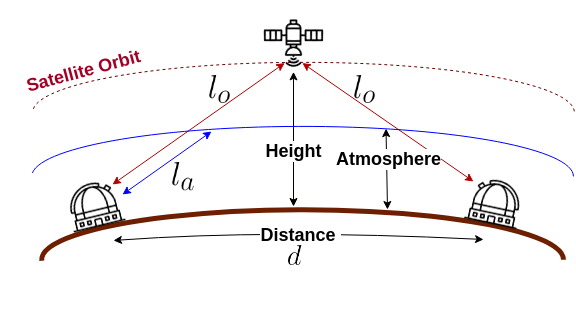}
    \vspace{-15pt}
    \caption{Satellite setup for generating EPR pairs over long distances.}
    \label{fig:satellite}
    \vspace{-10pt}
\end{figure}

Figure~\ref{fig:satellite} illustrates our satellite setup.
We assume two ground stations on Earth are separated by distance $d$ on the order of hundreds of kilometers. The satellite orbits at height $h$ and is equidistant from each ground station, at distance $l_o$. The satellite generates EPR pairs and sends half of each state toward each ground station. Photons travel a distance $l_o$ through free space of polynomially-decreasing transmissivity, and, once they reach the atmosphere, 
are subjected to a further decrease in transmissivity -- this time exponential with atmosphere attenuation coefficient $\alpha_a$ --   for the remaining distance to the ground station, $l_a$. 
In this setup, we consider optical links with circular apertures of diameters $d_s$ and $d_g$ for the satellite and ground station, respectively, that operate at wavelength $\lambda$. The upper bound for  transmissivity between the satellite and the ground station is approximated by
\begin{align}
\label{eq:satellite_transmissivity}
    \nonumber &\eta_o =\min((\pi d_s^2/4) (\pi d_g^2/4) /(\lambda l_o)^2,1),\\
    \nonumber &\eta_a = \exp(-\alpha_a  l_a),\\
    &\eta_s = \eta_o\eta_a,
\end{align}
where $\eta_o$ and $\eta_a$ are channel transmissivities corresponding to free space and the atmosphere, respectively, 
  and $\eta_s$ is the overall transmissivity ~\cite{paper:satellite_nitish,paper:saikat_satellite_loss_model}.

\section{Purification Protocols}
\label{section:methodology}
In this section, we first discuss the traditional way of carrying out purification via the pumping scheme.
Recall that we refer to this method
as the \textit{baseline protocol (BASE)}. We then introduce our \textit{optimistic protocol (OPT)}  and finally introduce the \textit{heralded-optimistic protocol (HOPT)} scheme briefly discussed in Sections~\ref{sec:intro}~and~\ref{section:related_work}. Throughout this section, we assume the time it takes for the EPR source to send out a new pair of photons to Alice and Bob is negligible compared to the propagation delay.

\begin{figure}[ht]
  \centering
  \subfigure[Example execution without failures.]{\includegraphics[width=0.8\linewidth]{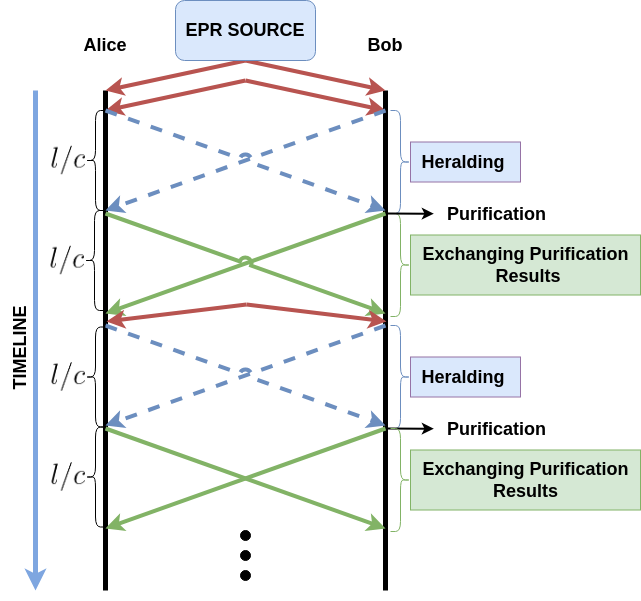}
  \label{fig:baseline_success}}\quad
  \subfigure[Example with EPR pair generation and purification failures.]{\includegraphics[width=0.8\linewidth]{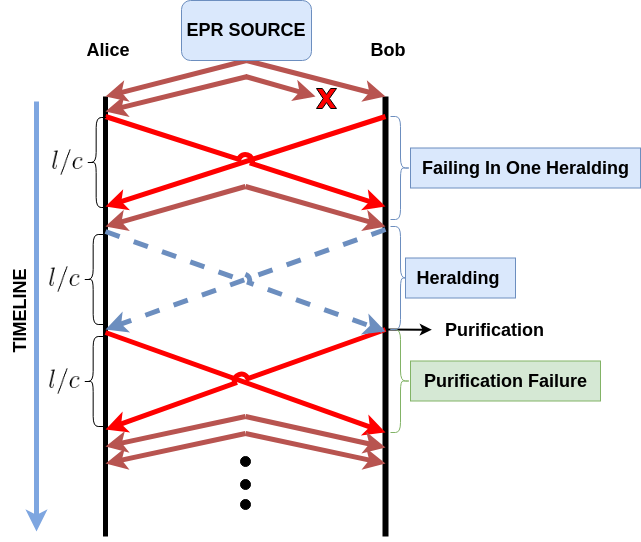}
  
  \label{fig:baseline_failure}}
  \caption{Baseline purification protocol event sequence and timing.}
    \label{fig:baseline}
\end{figure}
\begin{figure}[t]
  \centering
  \subfigure[Example execution without failures.]{\includegraphics[width=0.8\linewidth]{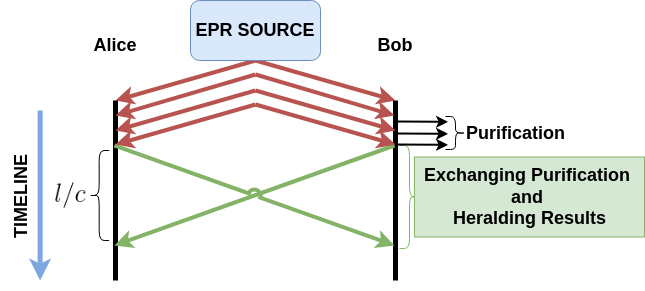}
  \label{fig:optimistic_success}}\quad
  \subfigure[Example with entanglement generation failure. Bob informs Alice, and the process restarts.]{\hspace{-2mm}\includegraphics[width=0.8\linewidth]{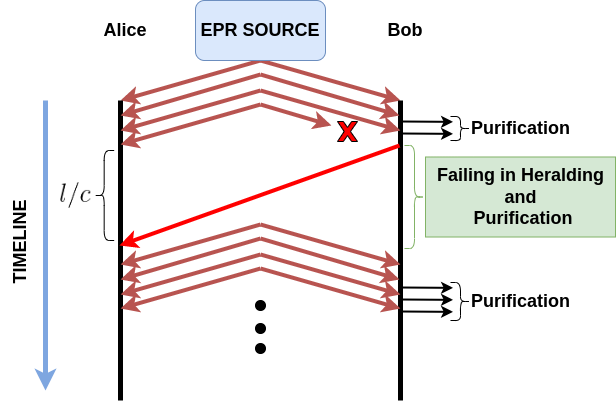}
    \label{fig:optimistic_failure}}
  \caption{Optimistic purification scheme event sequence and timing.}
    \label{fig:optimistic}
\end{figure}
\subsection{Baseline Protocol (BASE)}
In quantum networks,
heralding signals inform nodes of entanglement generation success or failure, and in case of the former, may also be used to provide information about necessary correction operations. Purification, being a probabilistic procedure, also requires classical information exchanges between participating nodes.
Figure~\ref{fig:baseline} exemplifies the sequence and timing of events for \textit{baseline} entanglement pumping. Figure~\ref{fig:baseline_success} depicts a successful purification procedure, while Figure~\ref{fig:baseline_failure} depicts a purification scenario where failure occurs in heralding and later in purification. In Figure~\ref{fig:baseline}, nodes Alice and Bob, each equipped with two quantum memories, are separated by $l$ km. An EPR pair source in the middle of the link sends half of the pair to Alice and the other half to Bob. Alice and Bob both know the rate of the EPR source and have synchronized clocks that tell them when they should expect to receive photons. 
Thus, upon each clock tick, any party that has not received their portion of the EPR pair informs the other party of the failure.
At the beginning of protocol, once they receive two EPR pairs, they send heralding signals, which take at least $l/c$ seconds to transmit, where $c$ is the speed of light in optical fiber -- 200,000 km/s. Then they perform purification and send the results through a classical channel while waiting for the purification results.

In the following, we go through the timeline presented in Figure~\ref{fig:baseline}: The nodes $(i)$ receive their portion of the main entangled pair from the midpoint source, $(ii)$ receive their portion of an auxiliary (sacrificial) entangled pair, $(iii)$ herald entanglement generation success/failure.
These steps are repeated until both pairs are successfully received. Next, the nodes
$(iv)$ execute a set of quantum gates and measurements on both sides, and $(v)$ exchange measurement results via classical messages.
The nodes then compare results, and if purification succeeds, they repeat the process from step $(ii)$ until a desired number of purification steps is achieved.
If the results indicate that purification failed, the nodes discard the main pair and restart from step $(i)$. 
If on the other hand, either party receives no photon, then a failure signal is sent.
As soon as two EPR pairs are established, the nodes initiate purification
as outlined above.
\subsection{Optimistic Protocol (OPT)}

The main idea behind the optimistic protocol is to proceed with all purification steps without waiting for any heralding or consistency checks until the very end. Figure~\ref{fig:optimistic} illustrates the optimistic protocol timeline, where Alice and Bob are equipped with the same hardware as in the baseline setup. Panel~\ref{fig:optimistic_success} depicts a scenario where all entanglement generation attempts and purification steps are successful, while panel~\ref{fig:optimistic_failure}, presents a scenario in which Bob does not receive his portion of an EPR pair, and the procedure is restarted.

In the following, we go through the timeline of Figure~\ref{fig:optimistic}: $(i)$ Alice and Bob receive their portion of the main and sacrificial EPR pairs from a midpoint source and each node that receives her/his portion does not wait for the heralding signal and will continue to execute required quantum gates for purification. A node that does not receive one or both photons will inform the other party of failure, causing all pairs to be discarded, and the process to restart. Upon success, the nodes go to step $(ii)$, where they execute local quantum gates and measurements to carry out purification, then they exchange purification results but they do not wait to receive them from the other end. 
The nodes then $(iii)$ receive the next sacrificial EPR pair and perform another round of purification without waiting for any heralding signals or purification results. 
As previously, if a node detects entanglement generation failure or purification failure, it informs its partner, taking the process back to step $(i)$.
The nodes repeat step $(iii)$ until a desired number of purification steps are completed. Finally, the nodes $(iv)$ check the final purification measurement outcomes to verify whether the purification steps were successful, going back to step $(i)$ in case a failure occurs.  
Note that in a setup where the propagation delay is larger than the time takes for the EPR source to send out a new pair of photons to Alice and Bob, end nodes do not check the results at the very end of a purification procedure, but they check the purification/heralding results in time when they are waiting to receive a new EPR pair and determine the next action based on the purification/heralding results.
\subsection{Heralded Optimistic Protocol (HOPT)}
We now introduce the \textit{heralded-optimistic protocol (HOPT)} which lies between the optimistic and baseline approaches. In this protocol, Alice and Bob wait only for each others' heralding signals -- they are optimistic about purification results and exchange them only at the very end of the process. This protocol was first introduced by Hartmann \emph{et al.}~\cite{paper:blind_purification}, and its distance scalability was later studied in~\cite{paper:blind_purificaiton_2nd}.

In this work, we modify the original HOPT protocol such that end-nodes do not wait until the end of the whole purification procedure to exchange the purification results, instead they can exchange the purification results as soon as they measure their qubits. This modification allows Alice and Bob to be informed of purification failures earlier than the original protocol~\cite{paper:blind_purification} that checks at the very end, thereby preventing them from wasting EPR pairs on a failed purification. Because of early notice of purification failure, this modification improves the overall rate compared to the original protocol. We compare our proposed OPT with the improved HOPT.

\section{Evaluation}
\label{section:evaluation}

In this section, we compare OPT to BASE, HOPT, and sharing EPR pairs with no purification (NOP). We begin with the pumping scheme of~\cite{paper:pumping} for ground- and satellite-based settings, then continue with current state-of-the-art purification circuits of~\cite{paper:purification_last_stefan} for ground-based EPR generation. 
For entanglement pumping, we calculate the average rate and fidelity as a function of the total number of purification steps for a fixed hardware parameter set. Additionally, we examine the effect of different memory coherence times and EPR source rates on fidelity. 
We evaluate the circuit of~\cite{paper:purification_last_stefan} in a similar manner.
Last, we evaluate the QKD performance of different protocols for ground- and satellite-based EPR generation schemes.
To do so, we calculate the SKR for ground- and satellite-based setups for all protocols and show that, the optimistic protocol yields the highest SKR when memory coherence times and initial fidelities are low.

\begin{figure}
\newcommand\x{0.19}
  \centering
  \subfigure[Fidelity for different protocols]{\includegraphics[scale=\x]{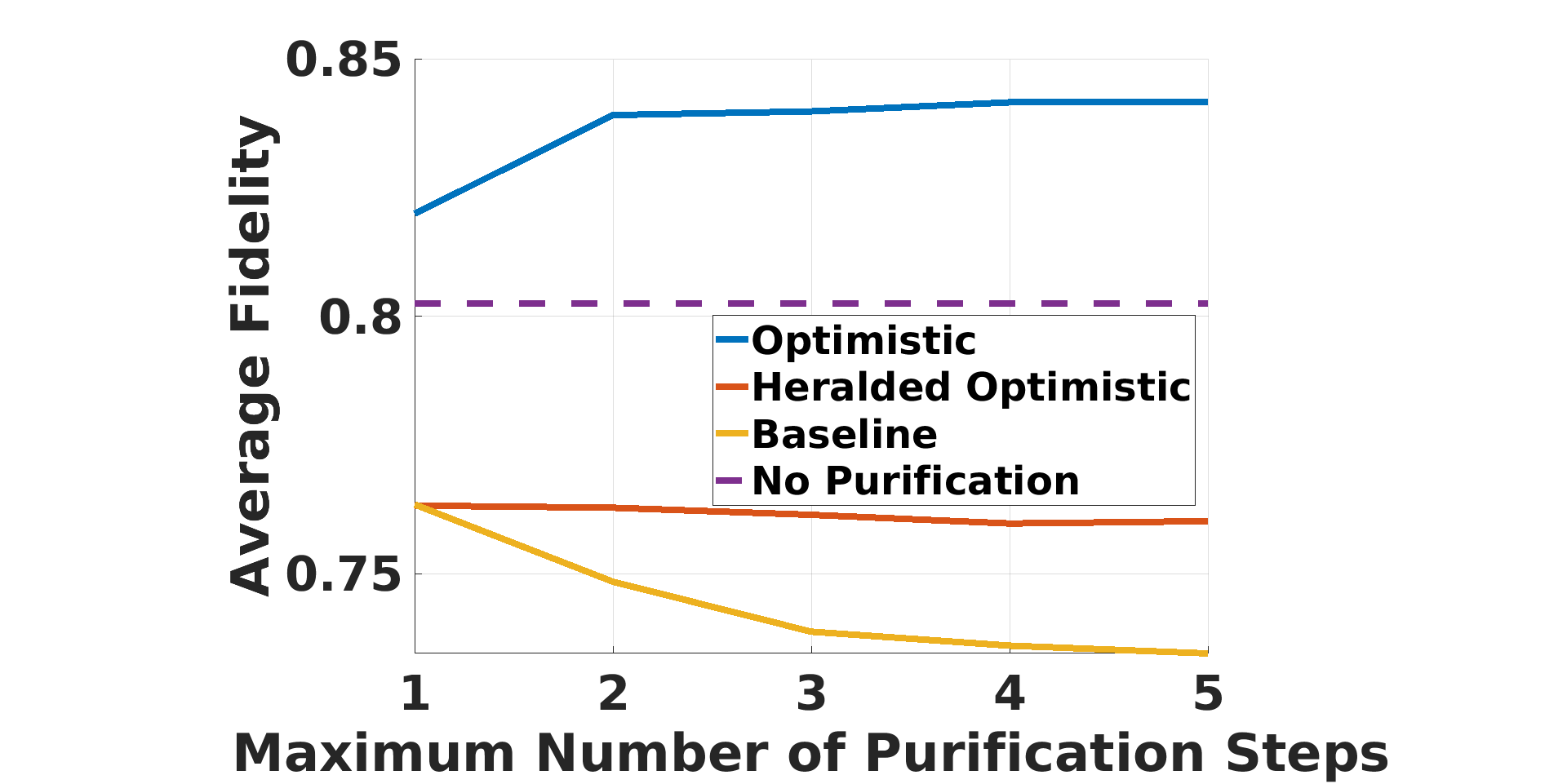}
  \label{fig:fid_pumping}}\quad
  \subfigure[Rate for different protocols]{\includegraphics[scale=\x]{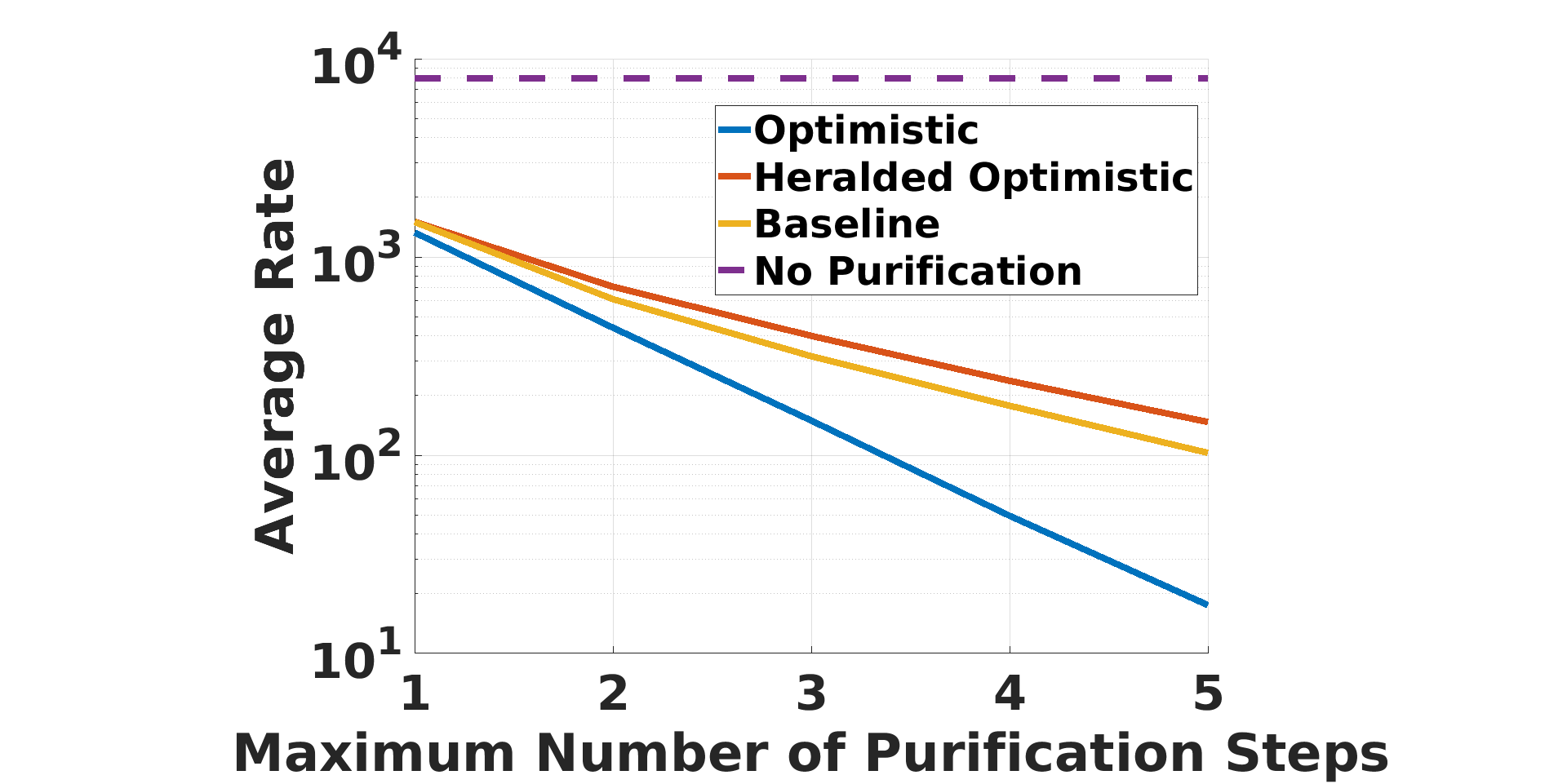}
    \label{fig:rate_pumping}}\quad
\caption{Fidelity and rate as functions of purification steps for different purification protocols and direct sharing with no purification, implementing entanglement pumping in ground-based setup. For the EPR source rate ($\mu$) of 1 GHz, the distance ($d$) of 20 km, and the initial fidelity ($F_0$) of $0.9$.}
\label{fig:fidelity_rate_pumping}
\end{figure}

\begin{figure}[t]
\centering
\includegraphics[scale=0.19]{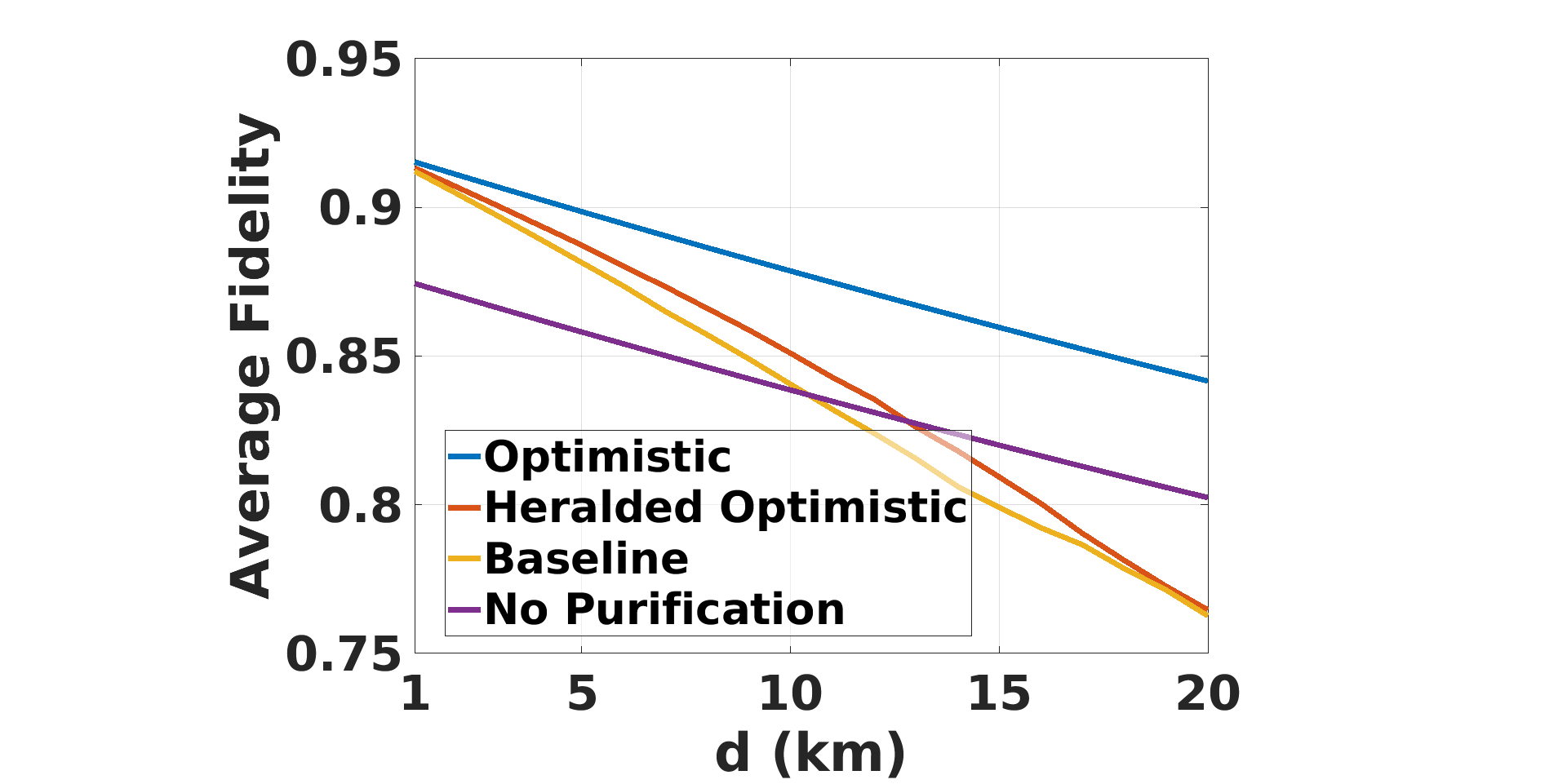}
\caption{Fidelity comparison for different protocols for distances ($d$) from 1 km up to 20 km in ground-based setup. For each data point, we plot the highest fidelity over five successive purification steps. EPR source rate ($\mu$), initial fidelity ($F_0$), and memory coherence time ($T_2$) are set to 1 GHz, 0.9, and 1 ms, respectively.}
\label{fig:fid_vs_dist}
\end{figure}

\begin{figure}[t]
\centering
\newcommand\x{0.19}
\hspace{-7mm}\includegraphics[scale=\x]{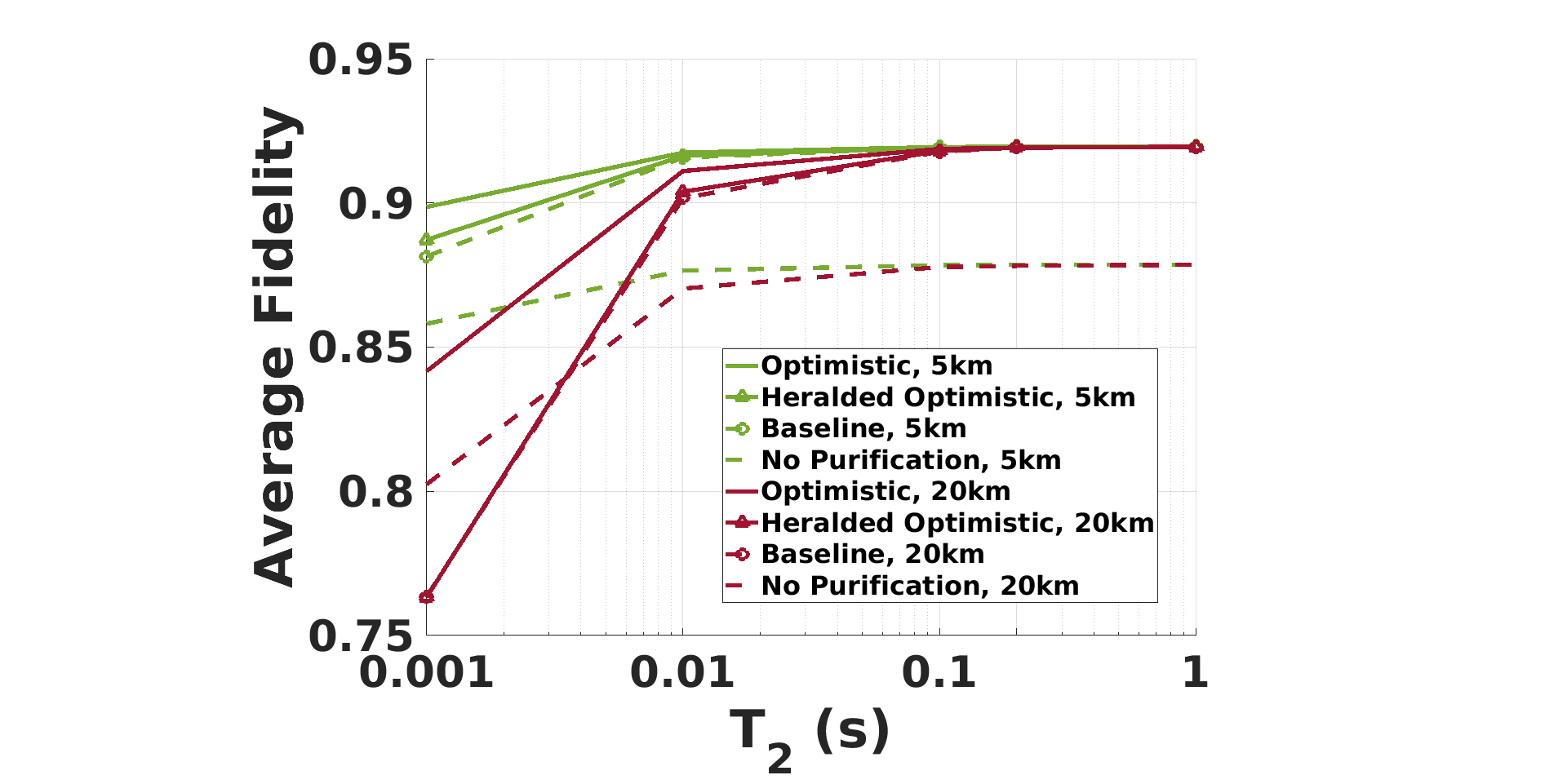}
\caption{The effect of memory coherence time ($T_2$) on average fidelity for entanglement pumping in ground-based setup. Initial fidelity ($F_0$) and EPR source rate ($\mu$) are set to 0.9 and 1 GHz, respectively. Here, all schemes' fidelities converge for $T_2\geq 0.1$ s. }
\label{fig:varying_t2_dejmps}
\end{figure}
\begin{figure}
\centering
\newcommand\x{0.19}
\includegraphics[scale=\x]{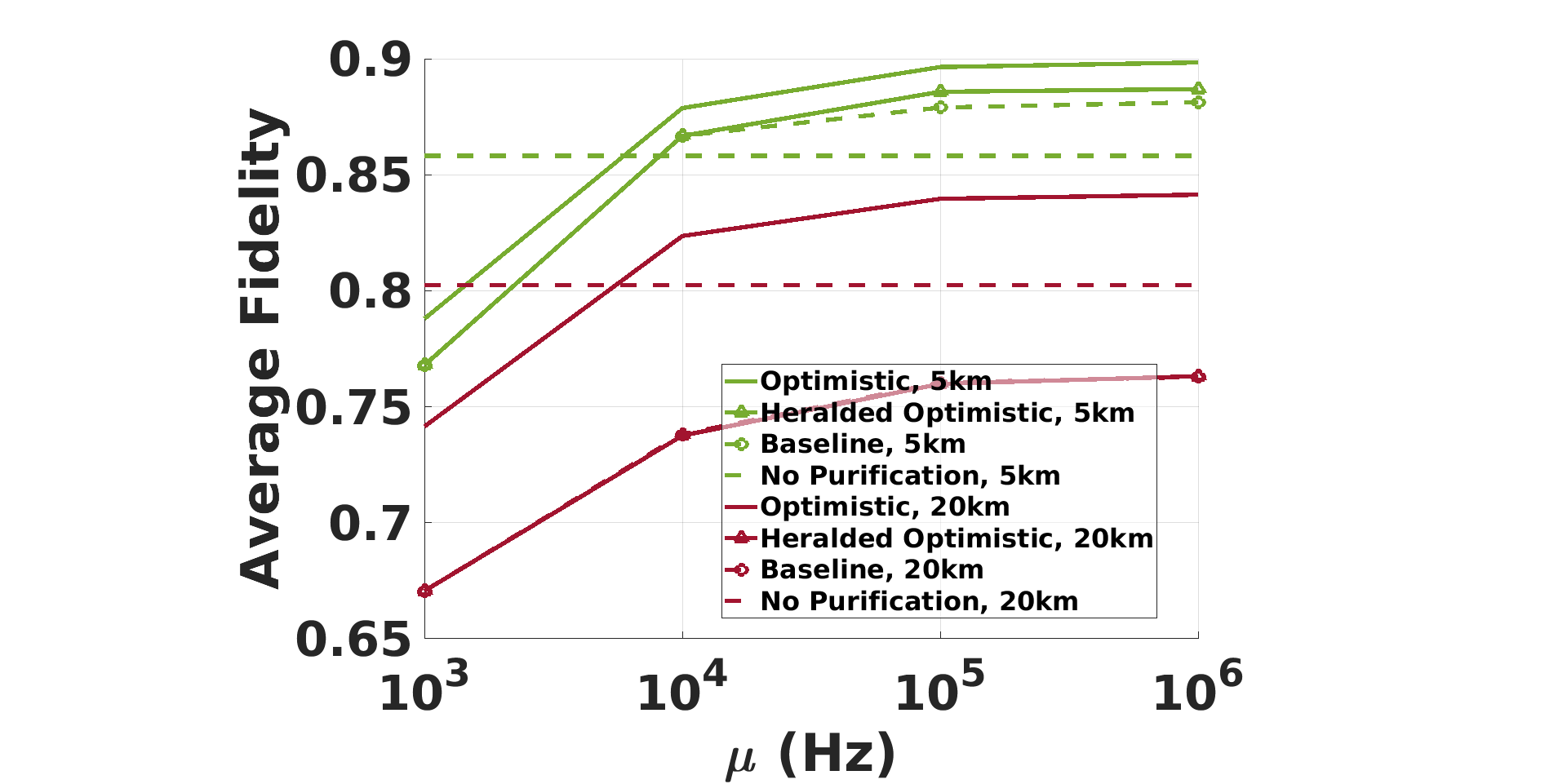}
\caption{The effect of EPR source rate ($\mu$) on the average output fidelity in ground-based setup. Initial fidelity ($F_0$) and memory coherence time ($T_2$) are set to 0.9 and 1 ms, respectively. Average fidelity increases with rate; the improvement becomes negligible beyond $10^6$ Hz. }
\vspace{-8pt}
\label{fig:varying_rate_dejmps}
\end{figure}
\subsection{Simulation Setup}
We evaluate each protocol for different combinations of memory coherence time ($T_2$), distance ($d$) between two nodes, initial fidelity ($F_0$), and EPR  source rate ($\mu$). For all cases, we utilize the Monte Carlo method. 
In our simulations, one simulation iteration starts with no shared entanglement and ends when the protocol 
successfully purifies a state.

For each combination of values, we perform 10,000 iterations, except for the QKD evaluation on circuits from~\cite{paper:purification_last_stefan} where we perform 50,000 iterations. Using these simulations, we calculate average fidelity and average rate of resulting entanglement, as well as the average SKR. For each simulation, we ascertain that the confidence interval is less than three percent of the average value. For all simulations, the noise parameters for gates, measurements, and memories are the same. For controlled gates, we assume depolarization with parameter $p_g = 0.99$, as per~\cite{paper:purification_last_stefan} (see~\eqref{eq:2qbit_depolarization}).  We assume imperfect measurement projection with parameter $p_m = 0.99$, as per~\cite{paper:purification_last_stefan} (see~\eqref{eq:imperfect_projection}). For memory noise, we assume amplitude damping ($T_1$) and dephasing ($T_2$) (see \eqref{eq:amplitudedamping} and~\eqref{eq:dephasing}). Since in our evaluation, we do not store qubits in memory for a long time (at most, in the regime of milliseconds) and $T_1$ for amplitude damping is typically on the order of minutes -- \textit{e.g.}, \cite{paper:linklayer} reports a $T_1$ of at least six minutes for Nitrogen-Vacancy (NV) center in diamond carbon atoms -- it is not a significant source of noise for a stored qubit. However, we include it in our simulation, setting $T_1$ to six minutes. On the other hand, $T_2$ is on the order of milliseconds, and up to seconds as observed in experiments~\cite{paper:linklayer,paper:nv_center_reviewer_suggestion_2}. For $T_2$, we evaluate our scheme from 0.001s up to 1s, increasing at a logarithmic scale. We set the fiber attenuation coefficient $\alpha_f$ to 0.2 as in~\cite{paper:netsquid}. We select initial fidelity $F_0$ from the range 0.75 to 0.90. $\mu$ is selected from the range 1 KHz to 1 GHz~\cite{paper:spooky, paper:earth_and_moon_rate}. Inter-node distance, $d$, varies from 1 km up to 20 km.

For the satellite setup, the distance between ground stations $d$ is at most 500 km and the satellite height is set to 400 km, matching the average altitude of the international space station~\cite{book:international_space_station}. The atmosphere extinction attenuation, $\alpha_a$, is set to 0.028125~\cite{paper:satellite_nitish}. Sender and receiver hardware parameters are set to a wavelength $\lambda = 737$ nm, a satellite optical link aperture $d_s = 0.2$ m, and a ground station optical link aperture $d_g = 2$ m~\cite{paper:satellite_nitish}.

\begin{figure}
\newcommand\x{0.19}
  \centering
  \subfigure[Fidelity for satellite-based setup]{\includegraphics[scale=\x]{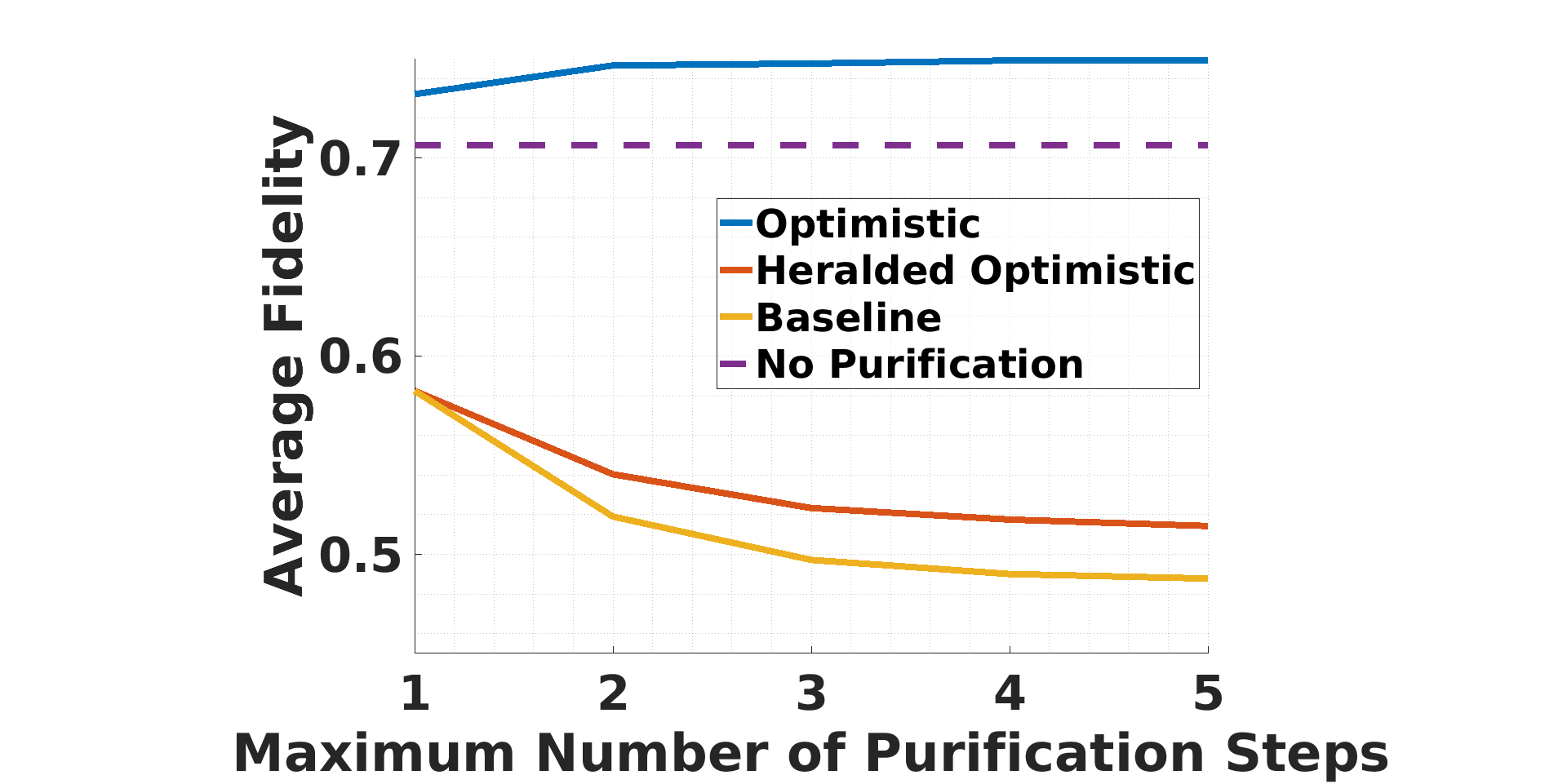}
  \label{fig:sat_fidelity}}\quad
  \subfigure[Rate for satellite-based setup]{\includegraphics[scale=\x]{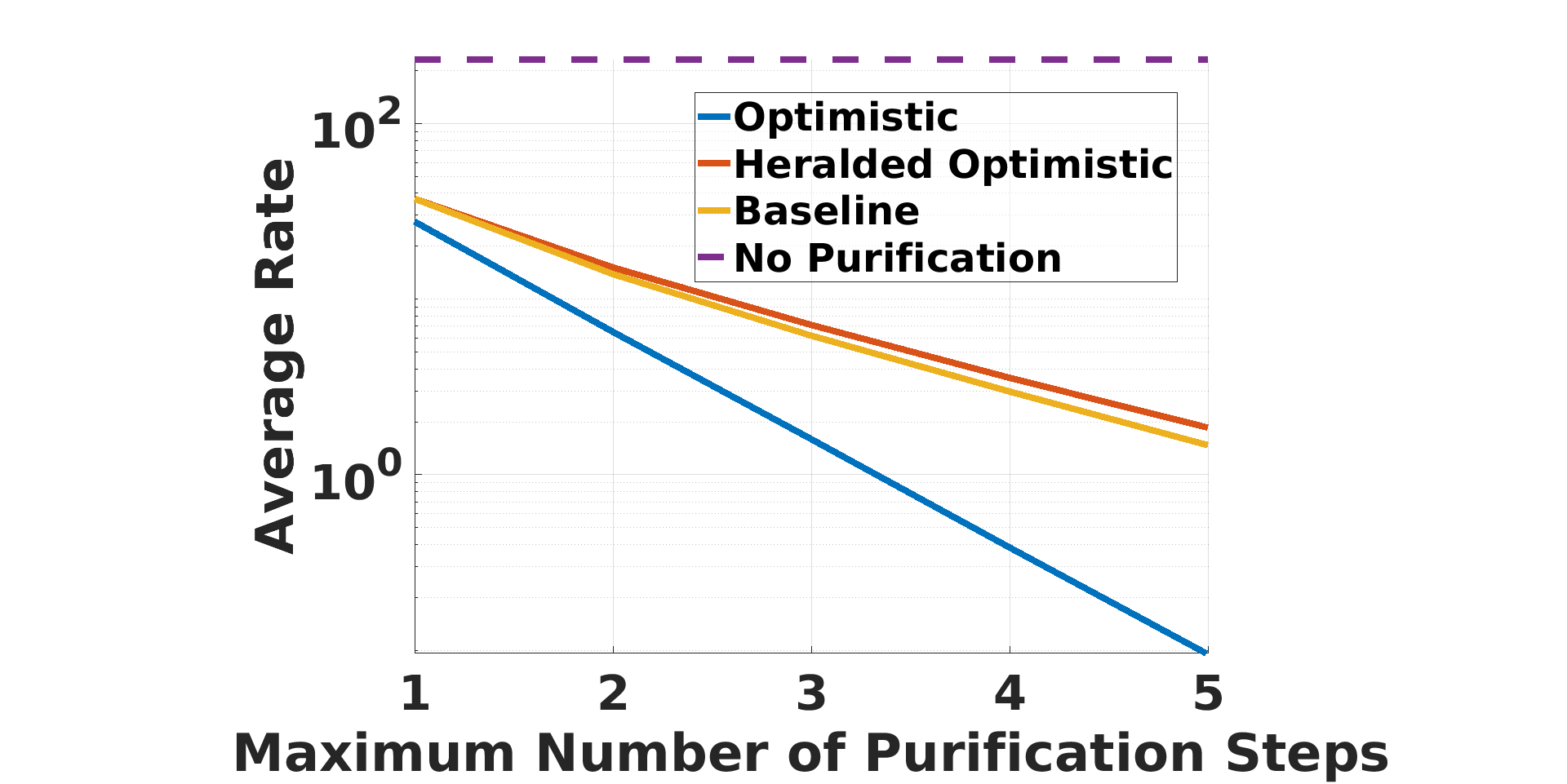}
    \label{fig:sat_rate}}\quad
\caption{Fidelity and rate comparison for different protocols implementing entanglement pumping, versus direct sharing without purification in satellite-based setup. Initial fidelity ($F_0$), EPR source rate ($\mu$), memory coherence time ($T_2$), and distance ($d$) are set to 0.9, 1 GHz, 10 ms, and 500 km, respectively.}
\label{fig:sat_fidelity_rate}
\end{figure}

\begin{figure}[t]
\includegraphics[scale=0.19]{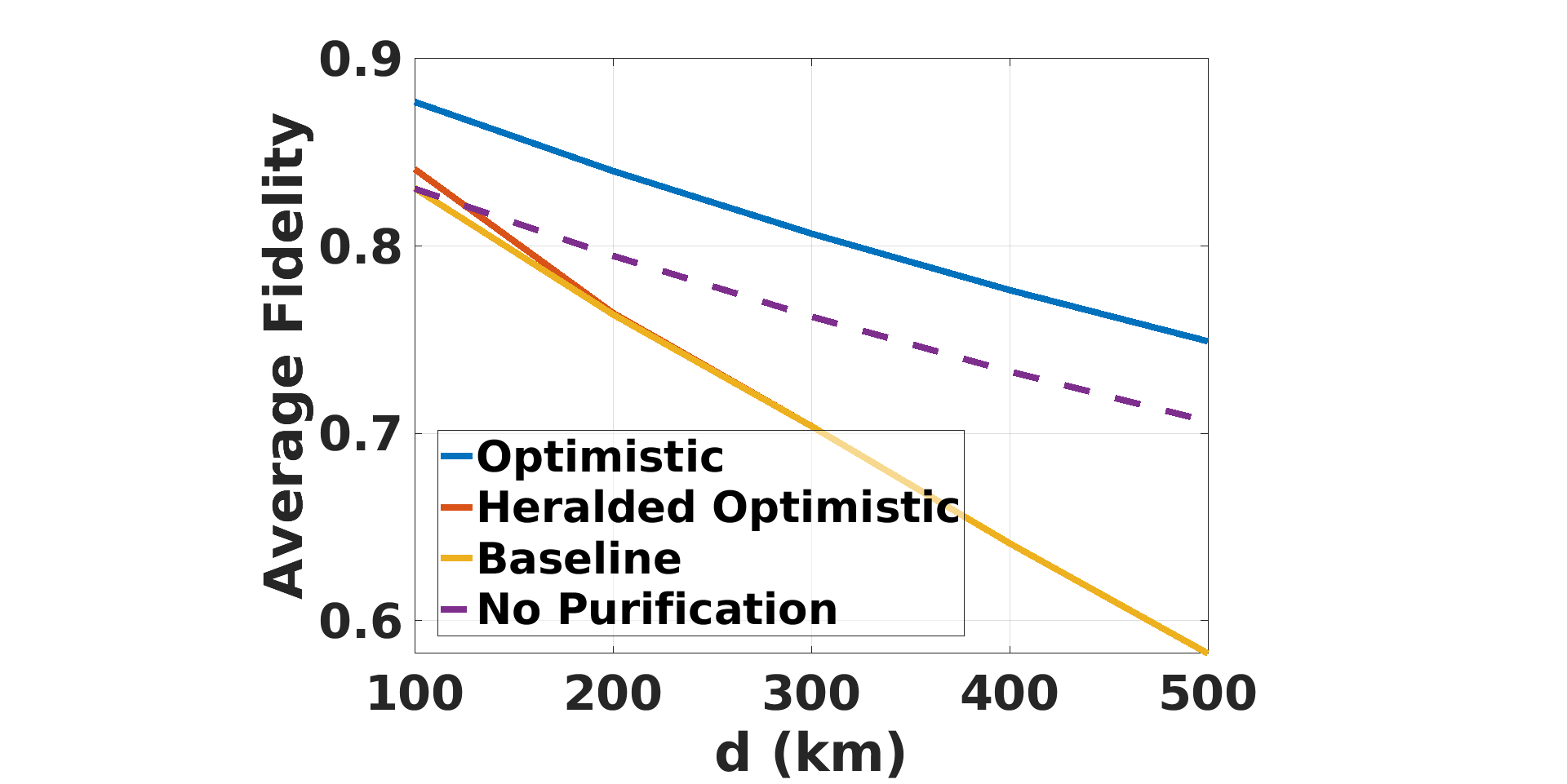}
\caption{Fidelity comparison for different protocols implementing entanglement pumping in satellite-based setup, for distance ($d$) from 100 km up to 500 km. For each data point of purification protocols, we plot the highest fidelity over five purification steps. We set initial fidelity ($F_0$) to 0.9, EPR source rate ($\mu$) to 1 GHz, and memory coherence time ($T_2$) to 10 ms.}
\label{fig:sat_fidelity_distance}
\end{figure}

\begin{figure}[t]
\includegraphics[scale=0.19]{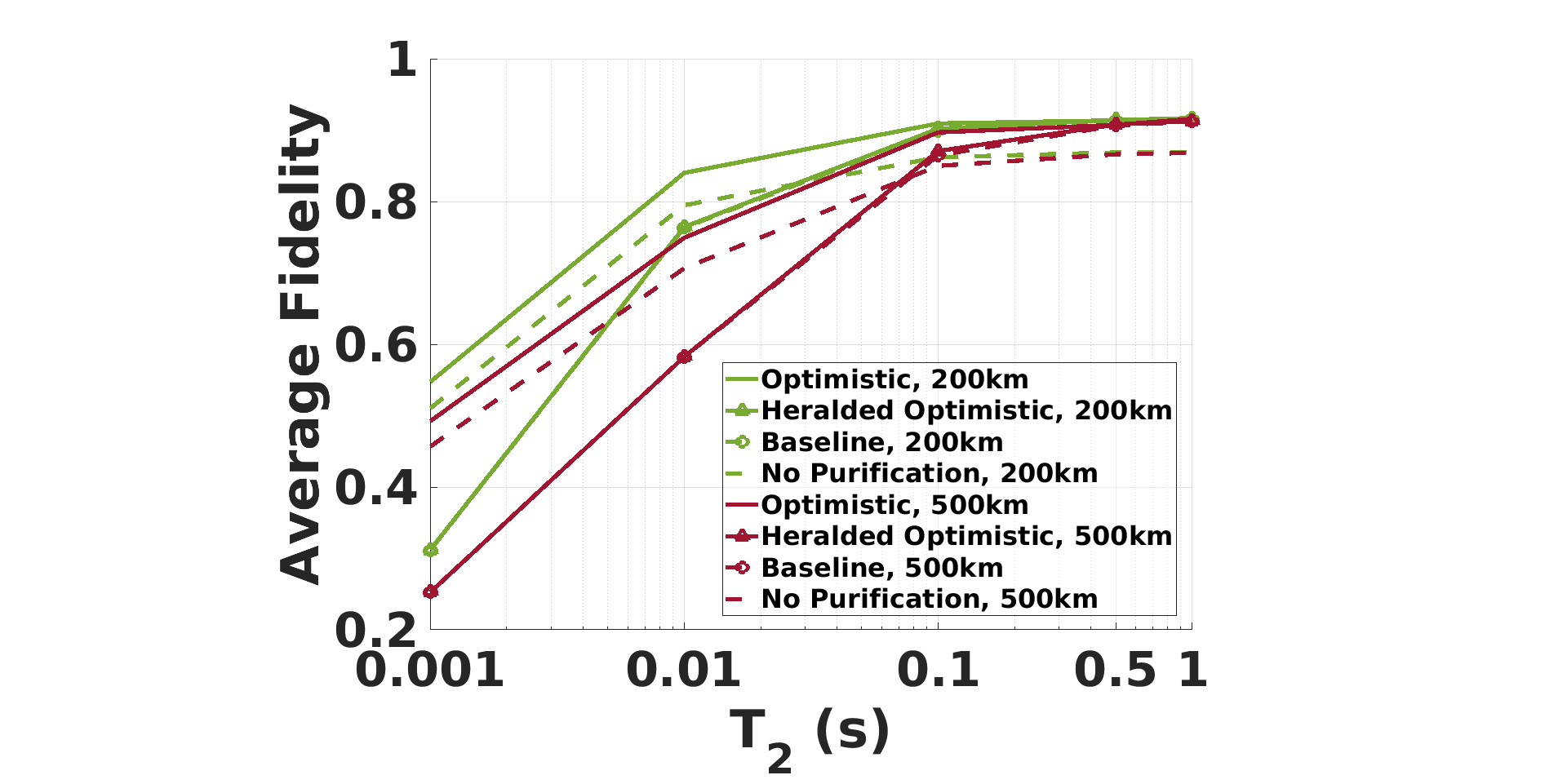}
\caption{The effect of memory coherence time ($T_2$) on the average fidelity in the satellite-based setup.  The initial fidelity ($F_0$) and the EPR source rate ($\mu$) are set to 0.9 and 1 GHz, respectively.
Increasing $T_2$ causes purification protocols to converge to the same fidelity.}
\label{fig:sat_varying_t2}
\end{figure}

\begin{figure}[t]
\includegraphics[scale=0.19]{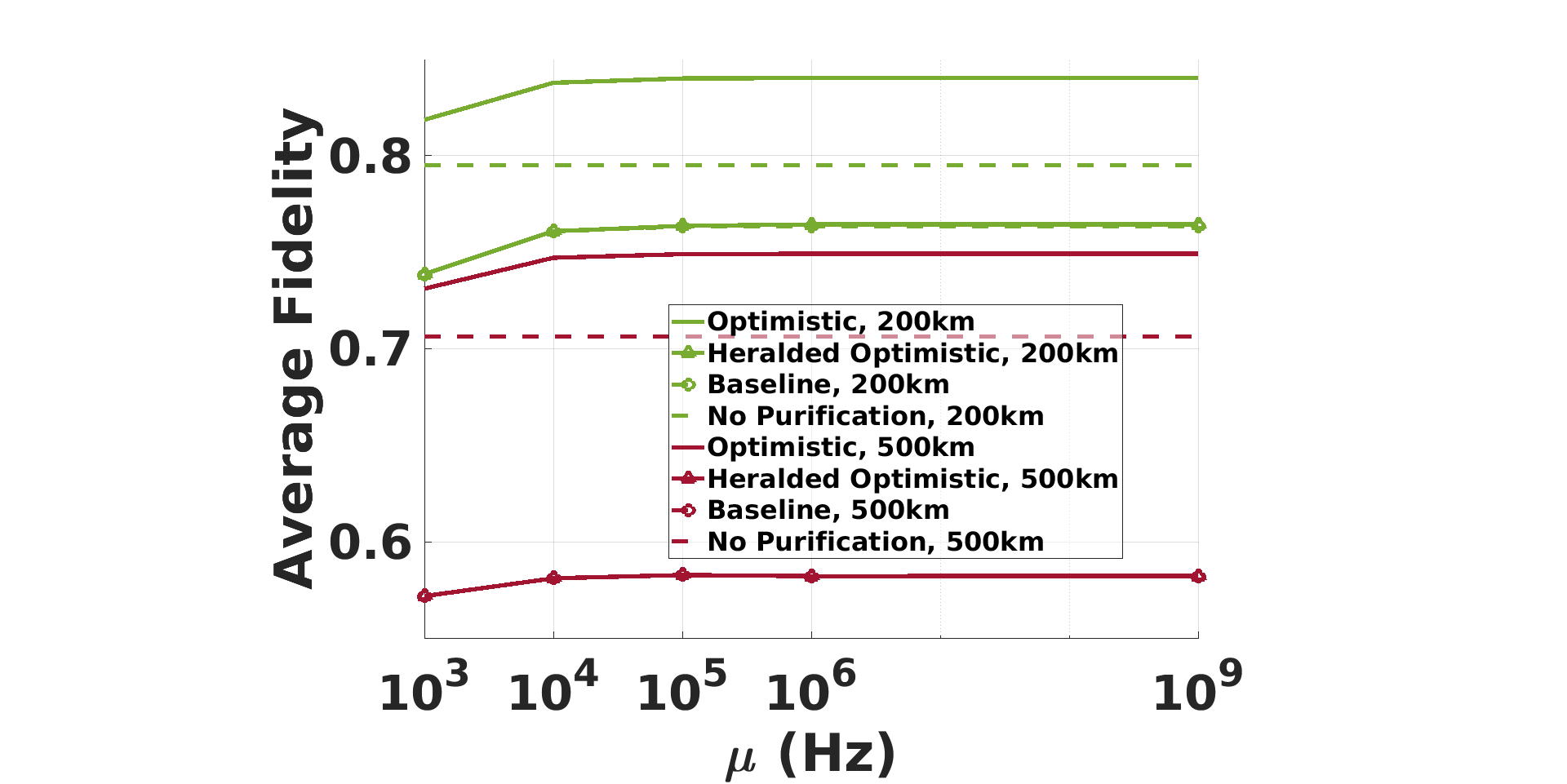}
\caption{The effect of EPR source rate ($\mu$) on the average fidelity in satellite-based setup for the initial fidelity ($F_0$) of 0.9 and the memory coherence time ($T_2$) of 10 ms.}
\label{fig:sat_varying_rate}
\end{figure}

\begin{figure}
\newcommand\x{0.19}
  \centering
  \subfigure[Fidelity for satellite-based setup]{\includegraphics[scale=\x]{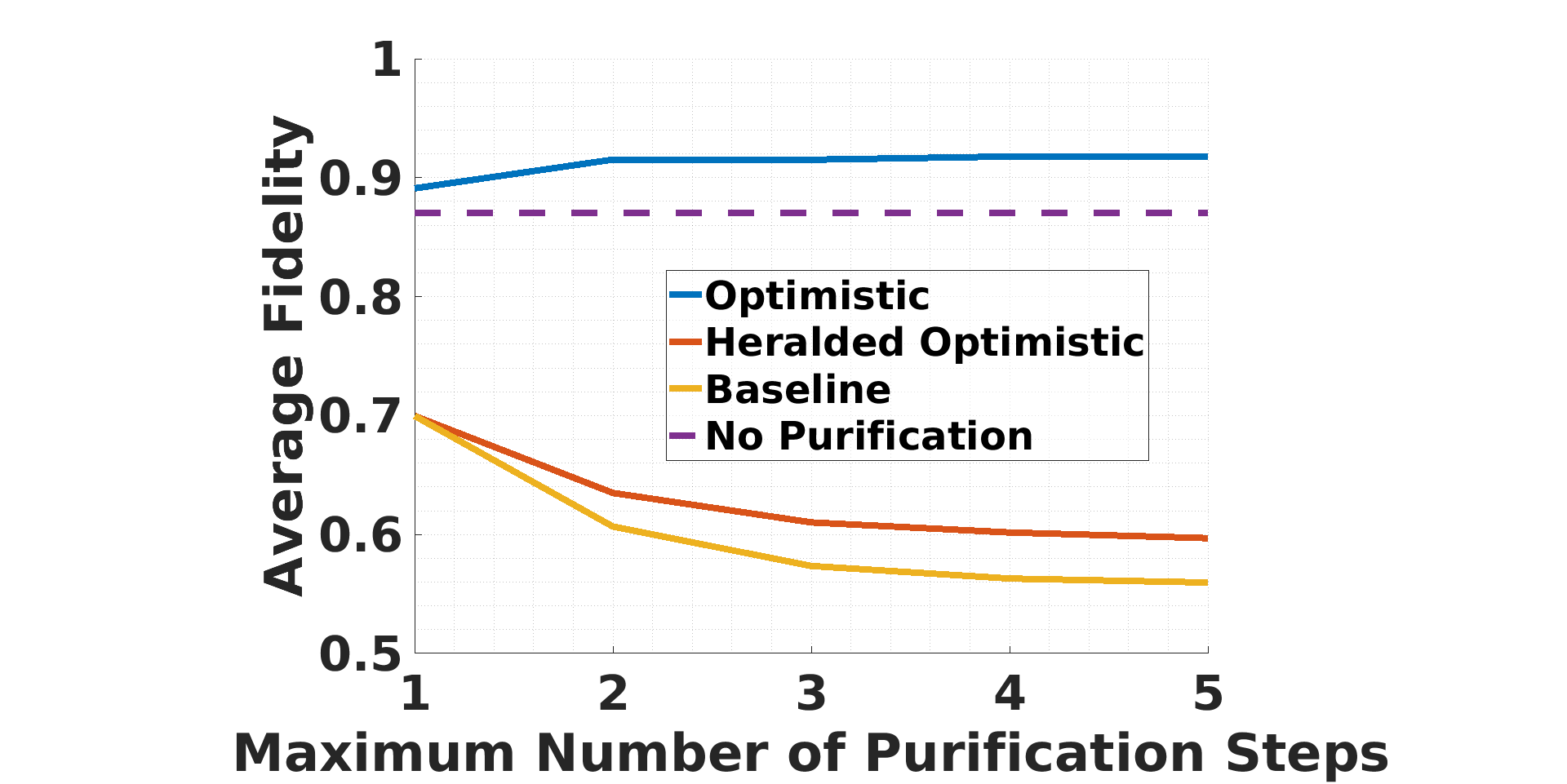}
  \label{fig:sat_fidelity_w/o}}\quad
  \subfigure[Rate for satellite-based setup]{\includegraphics[scale=\x]{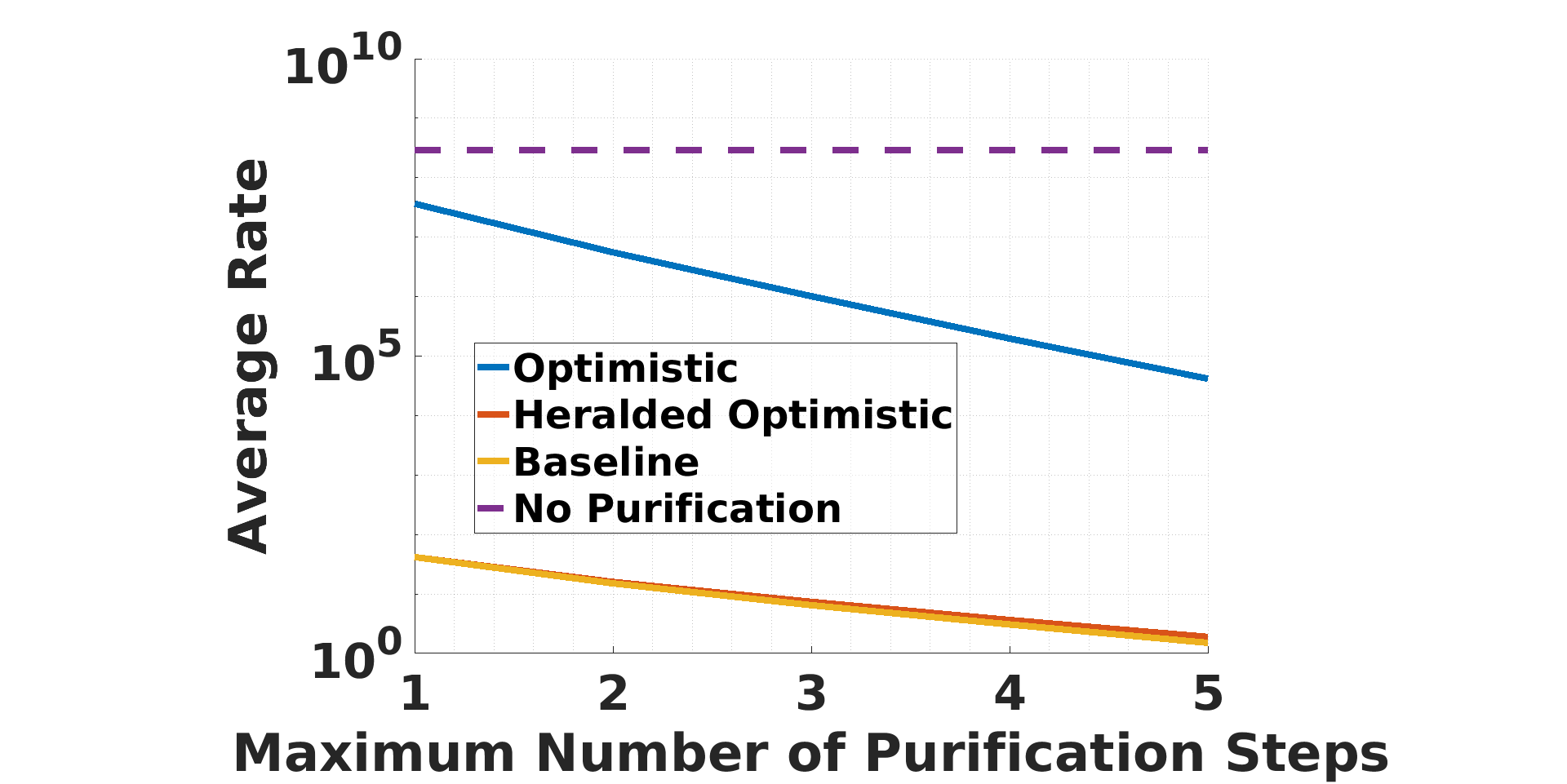}
    \label{fig:sat_rate_w/o}}\quad
\caption{Fidelity and rate comparison for different protocols without waiting for final confirmation in satellite-based setup. Initial fidelity ($F_0$), EPR source rate ($\mu$), memory coherence time ($T_2$), and distance ($d$) are set to 0.9, 1 GHz, 10 ms, and 500 km, respectively.}
\label{fig:sat_fidelity_rate_wo_confirmation}
\end{figure}

\begin{figure}
\newcommand\x{0.18}
  \centering
  \subfigure[Fidelity for optimized purification circuit.]{\includegraphics[scale=\x]{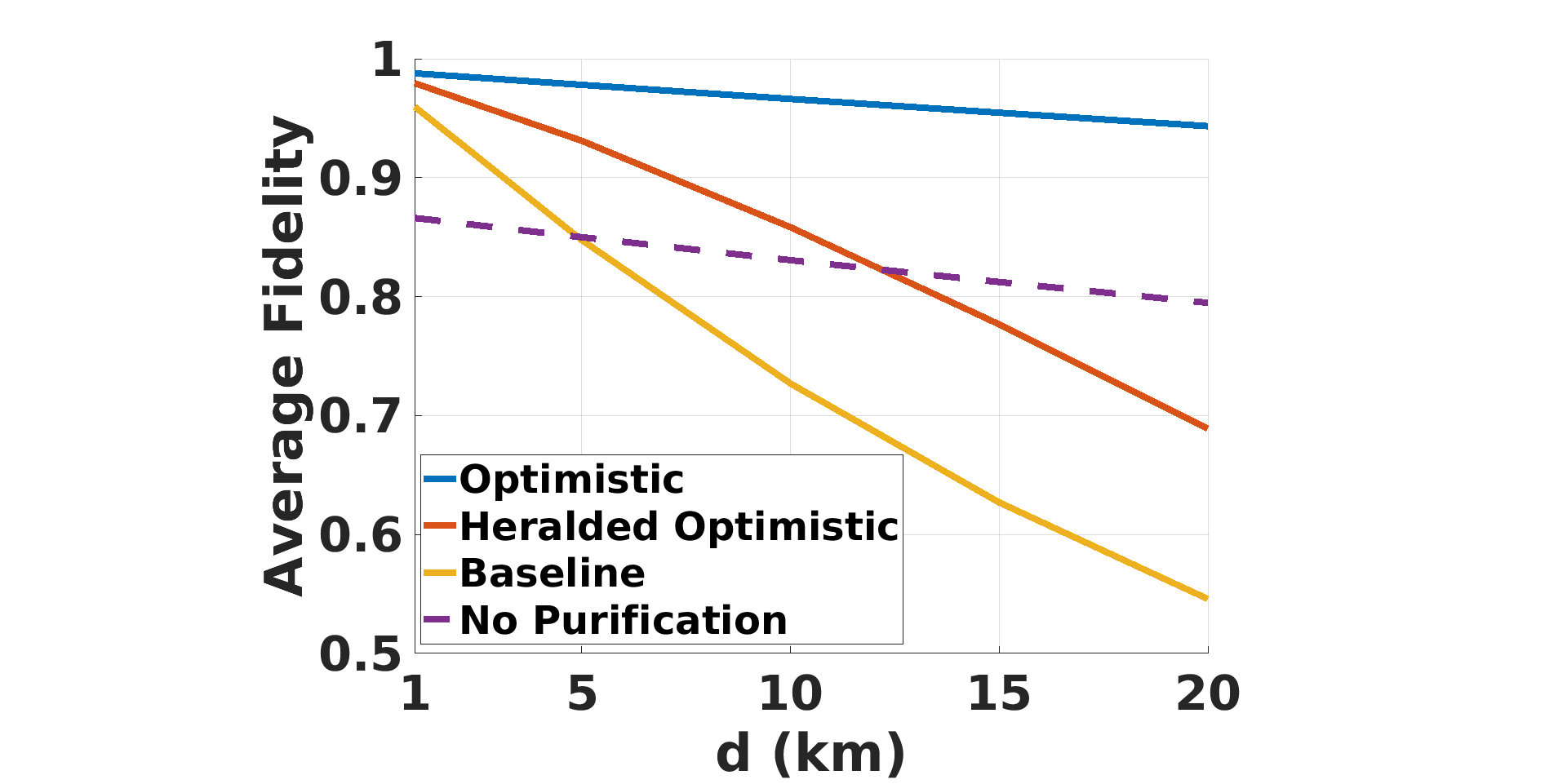}
  \label{fig:optimized_fidelity}}\quad
  \subfigure[EPR rate for optimized purification.]{\includegraphics[scale=\x]{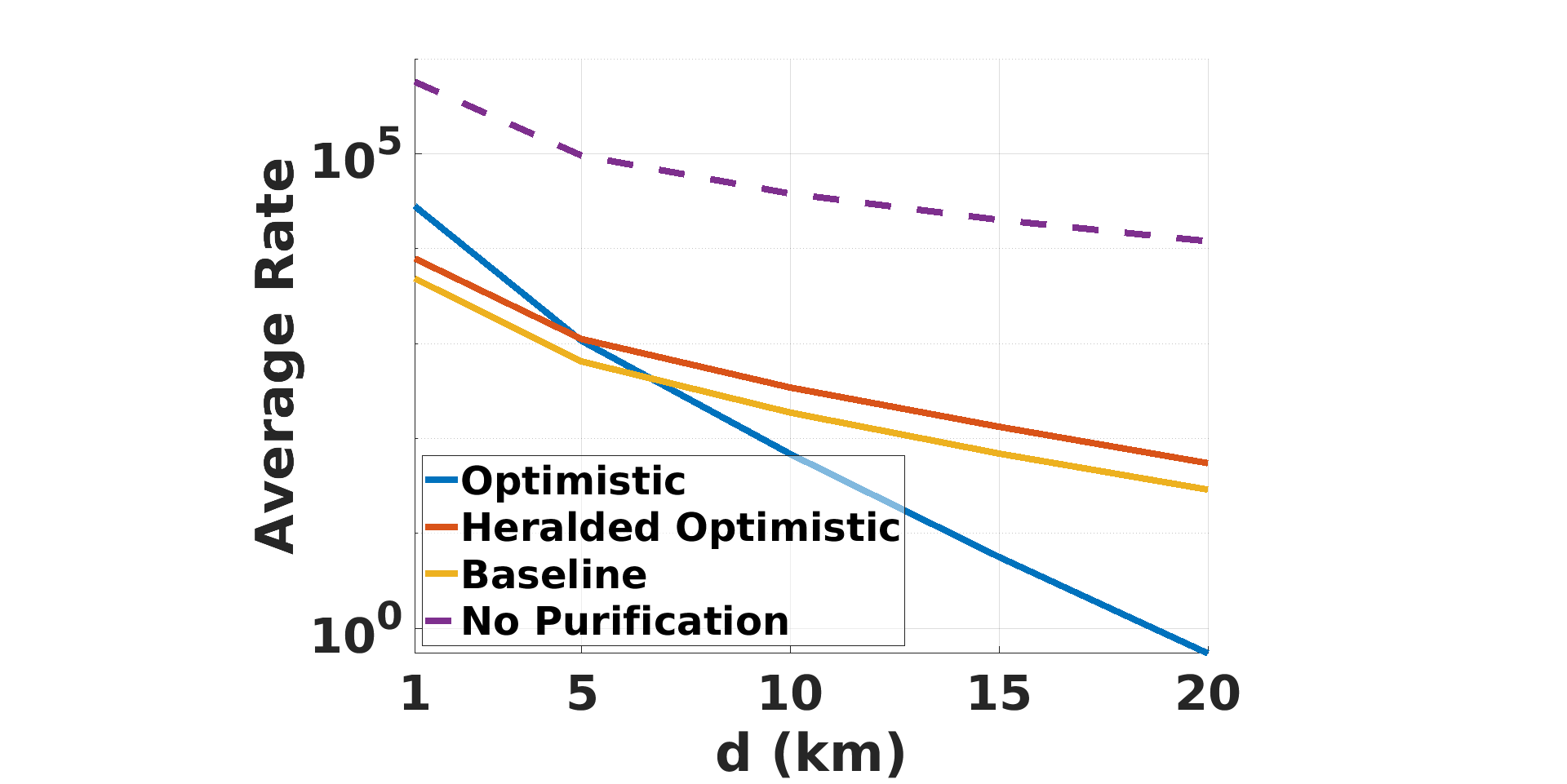}
    \label{fig:optimized_rate}}\quad
\caption{Final fidelity and rate for optimized purification circuit in ground-based setup. For initial fidelity ($F_0$) of 0.9, a memory coherence time ($T_2$) of 1ms, and an EPR source rate ($\mu$) of 1 GHz.}
\label{fig:optimized_circuit_rate_fidelity}
\end{figure}

\begin{figure}
\newcommand\x{0.19}
\centering
\includegraphics[scale=\x]{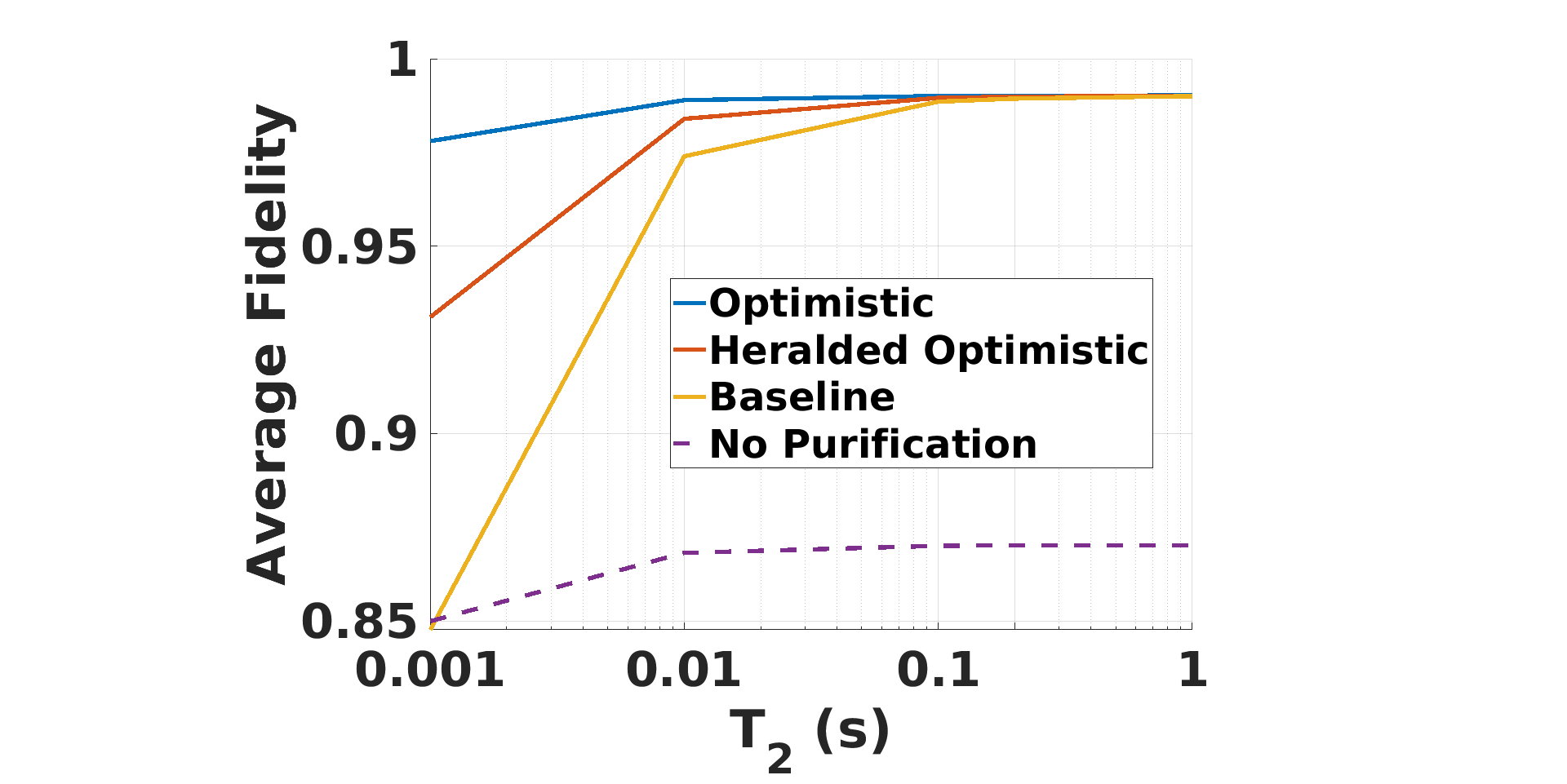}
\caption{The effect of memory coherence time ($T_2$) on the average fidelity in the optimized circuit for ground-based setup. Initial fidelity ($F_0$), distance ($d$), and EPR source rate ($\mu$) are set to 0.9, 5 km, and 1 GHz, respectively. Increasing $T_2$ causes protocols to converge to the same fidelity.}
\label{fig:varying_t2_optimized}
\end{figure}
\begin{figure}[h]
\centering
\newcommand\x{0.19}
\hspace{6mm}\includegraphics[scale=\x]{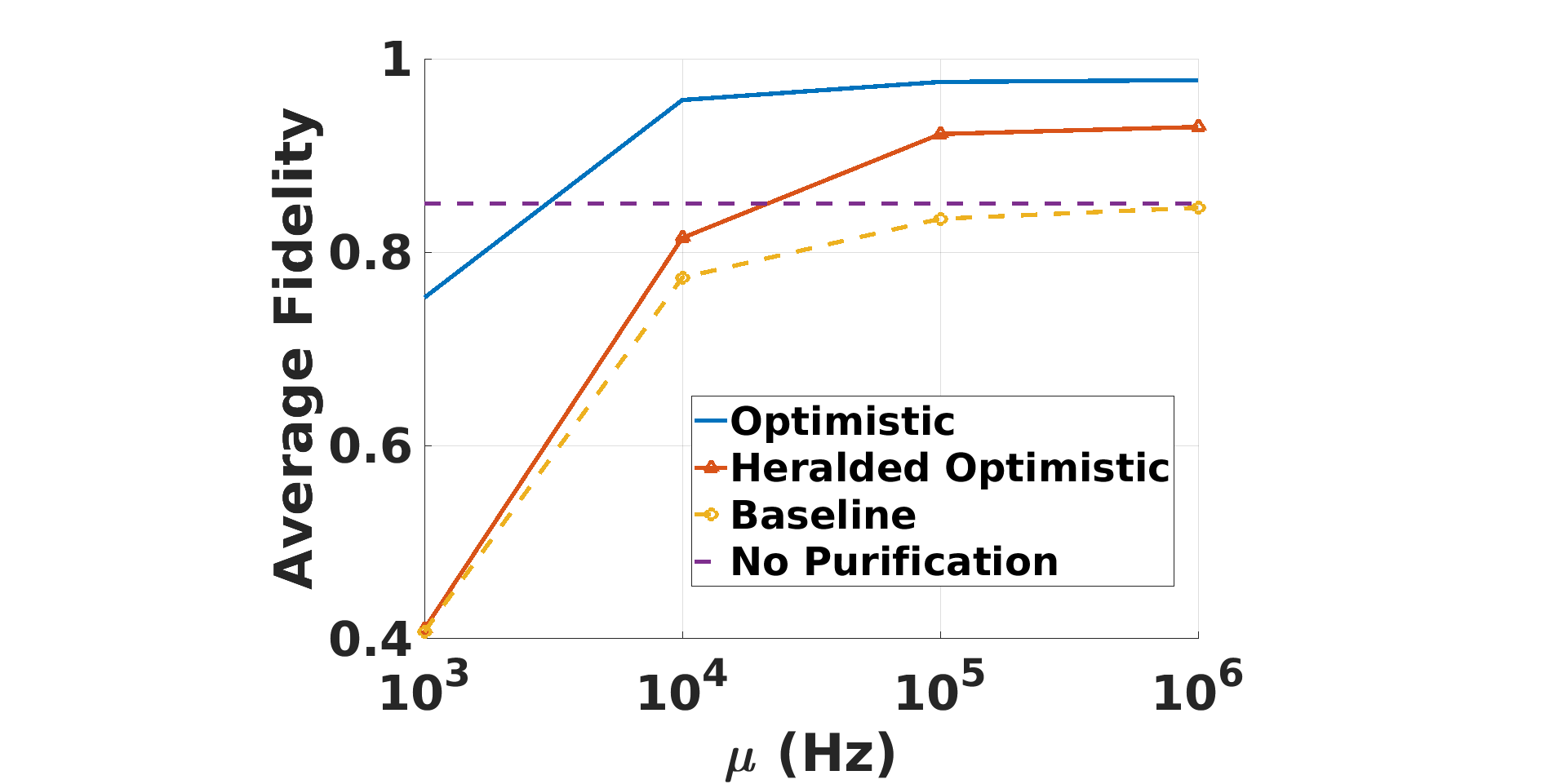}
\caption{The effect of source rate on output fidelity for the optimized circuit in ground-based setup. Initial fidelity ($F_0$), distance ($d$), and memory coherence time ($T_2$) are set to 0.9, 5 km, and 1 ms, respectively.}
\label{fig:varying_rate_optimized}
\end{figure}

\begin{figure*}[h]
\newcommand\x{0.15}
  \centering
  \subfigure[Ground-based setup with 1 KHz EPR source rate ($\mu$) and $d=20$ km.]{\includegraphics[scale = \x]{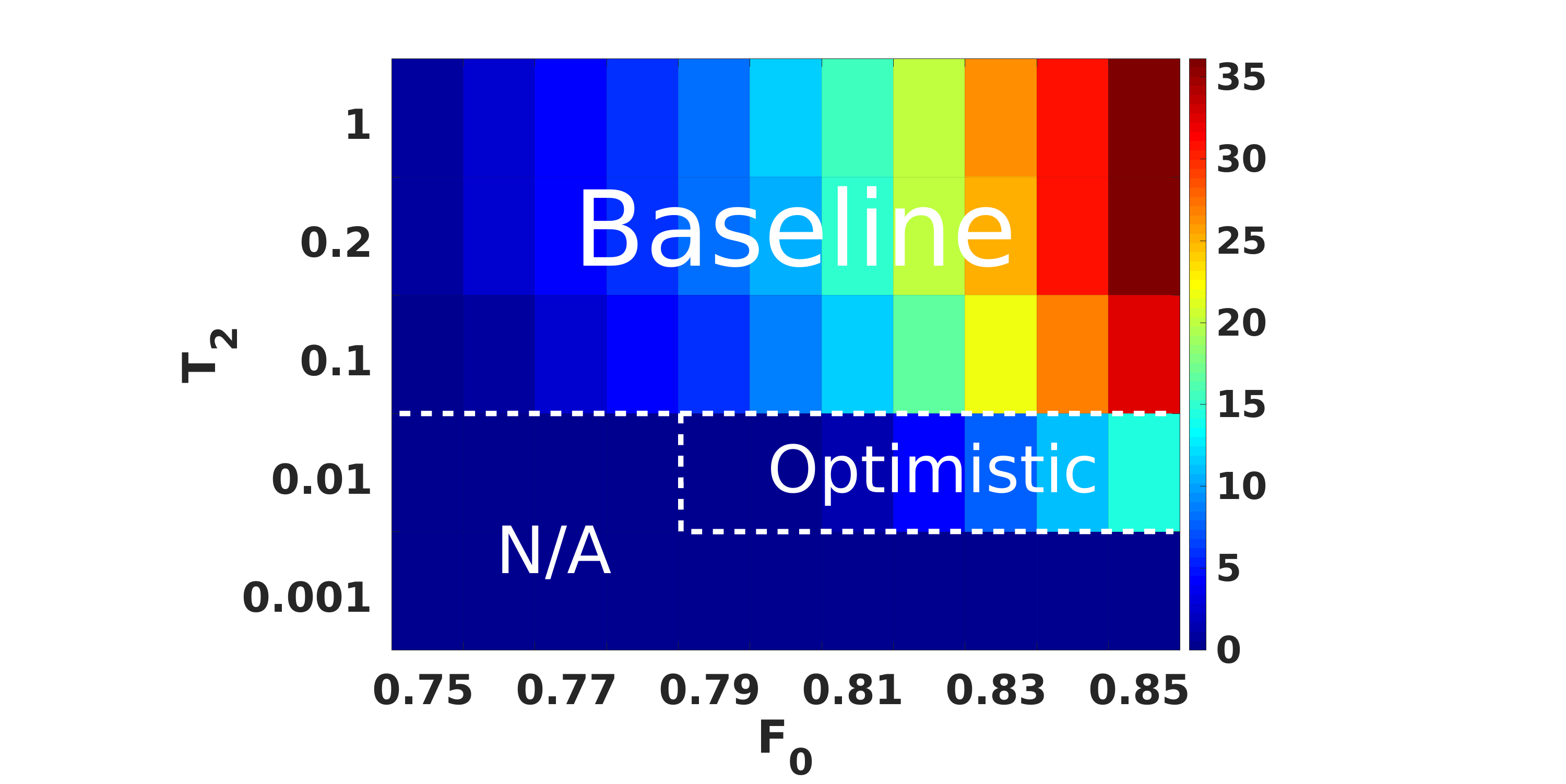}
  \label{fig:pumping_bb84_1}}\qquad
  \subfigure[Ground-based setup with 1 MHz EPR source rate ($\mu$) and $d=20$ km.]{\hspace{-4.5mm}\includegraphics[scale = \x]{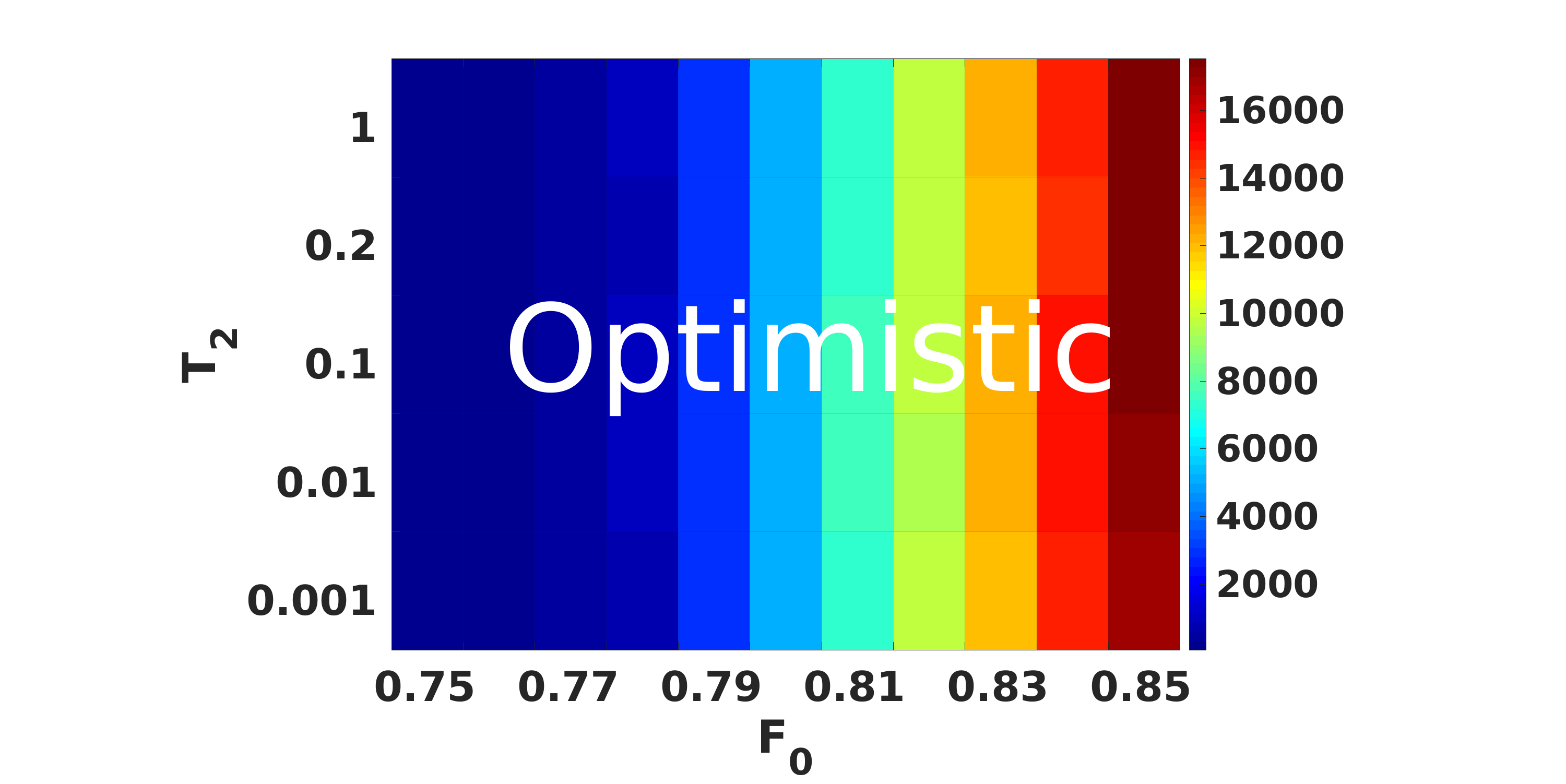}
    \label{fig:pumping_bb84_2}}\qquad
  \subfigure[Satellite-based setup with 1 KHz EPR source rate ($\mu$) and $d=500$ km.]{\hspace{1mm}\includegraphics[scale=\x]{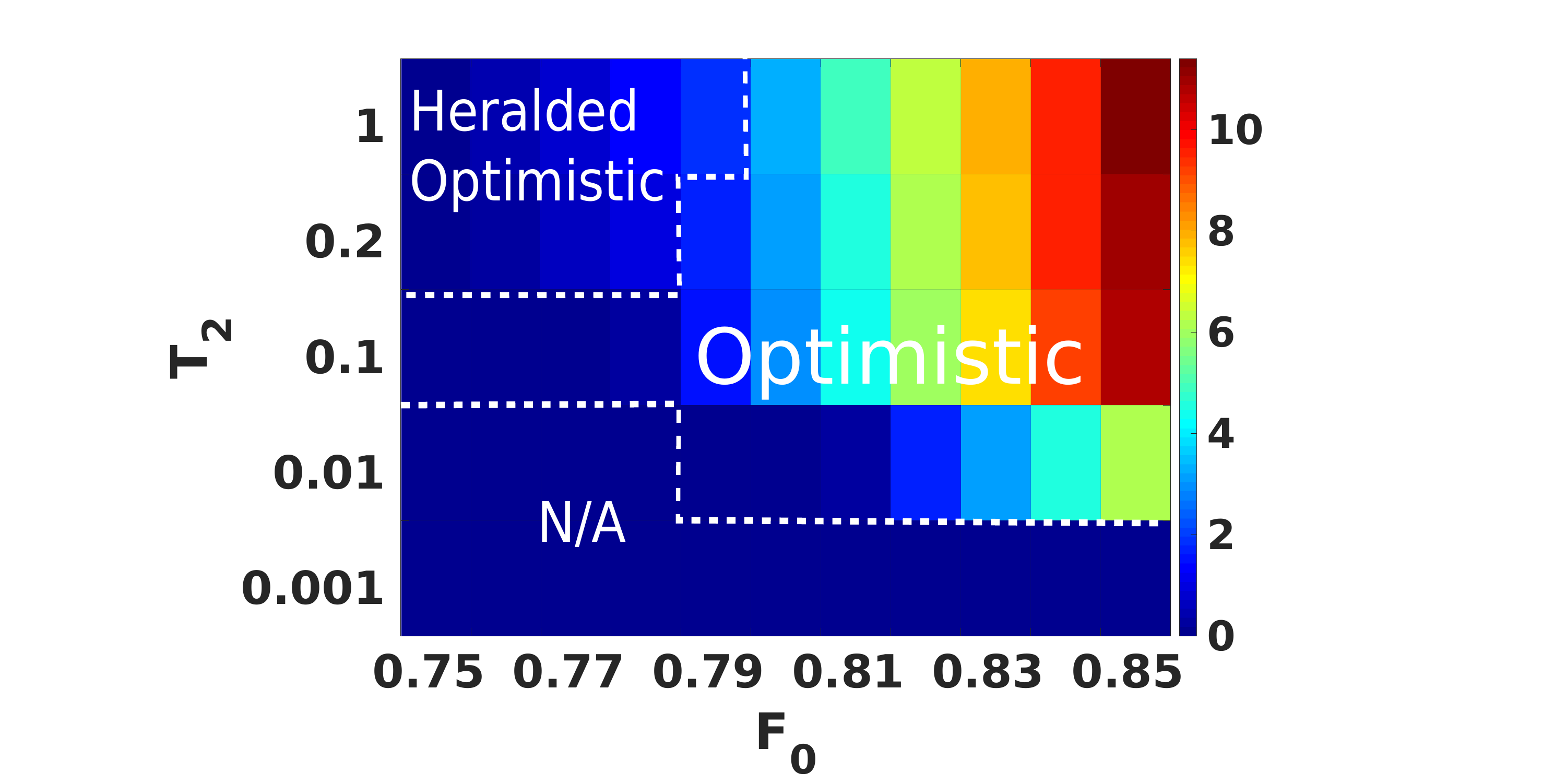}
    \label{fig:pumping_bb84_3}}\quad
  \subfigure[Satellite-based setup with 1 MHz EPR source rate ($\mu$) and $d=500$ km.]{\includegraphics[scale=\x]{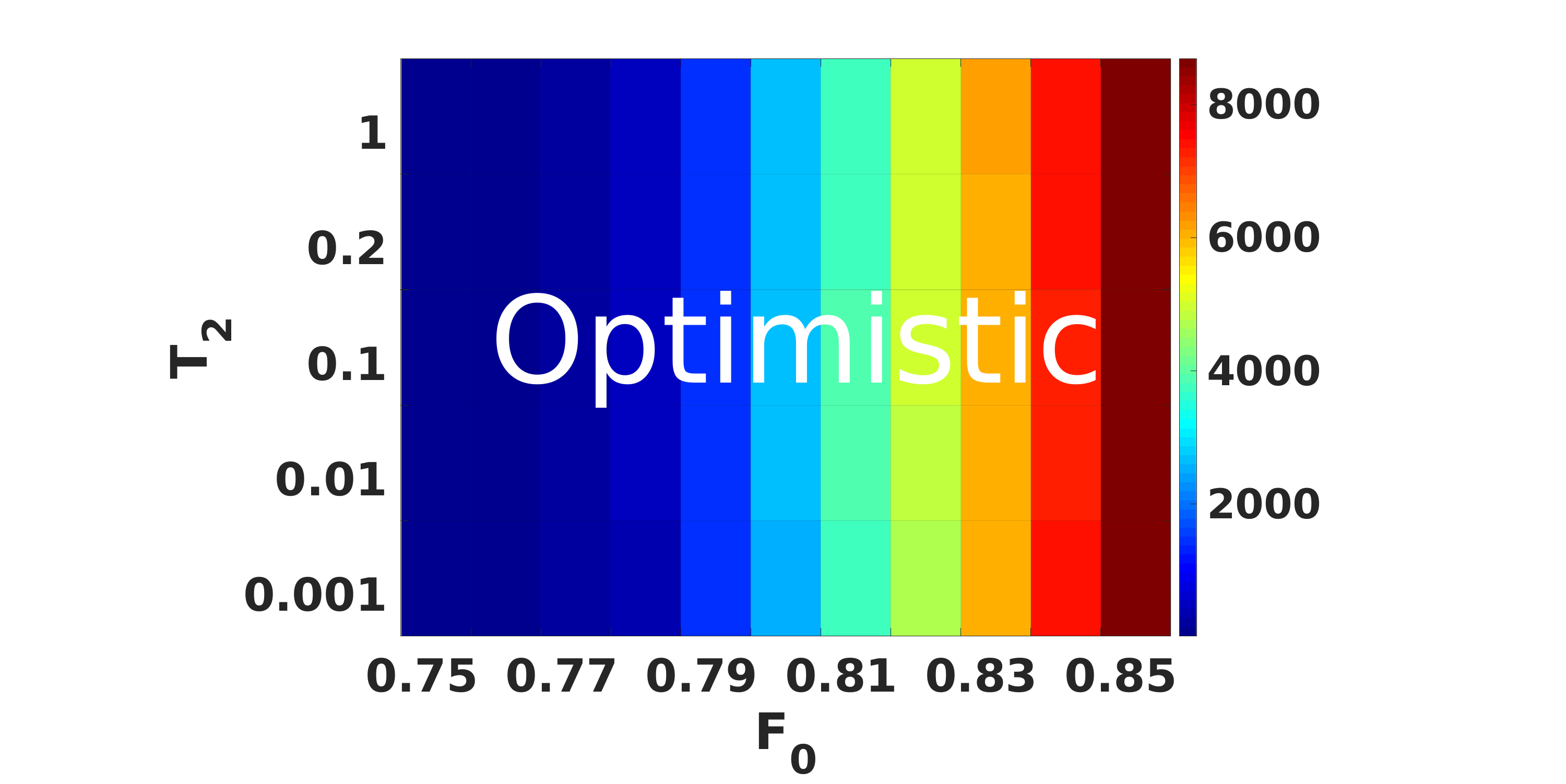}
    \label{fig:pumping_bb84_4}}
\caption{Heatmaps for BB84 SKR using the pumping scheme, as a function of $T_2$ and $F_0$ for various rates and EPR pair generation setups. We demarcate different regions with dashed lines and label each region to show the best protocol for QKD in that region. The `N/A' label indicates values of ($F_0$, $T_2$) where no positive SKR can be achieved.}
\vspace{-8pt}
\label{fig:pumping_bb84}
\end{figure*}
\begin{figure*}[h]
\newcommand\x{0.16}
  \centering
  \subfigure[Ground-based setup with 1 KHz EPR source rate ($\mu$) and $d=20$ km.]{\includegraphics[scale=\x]{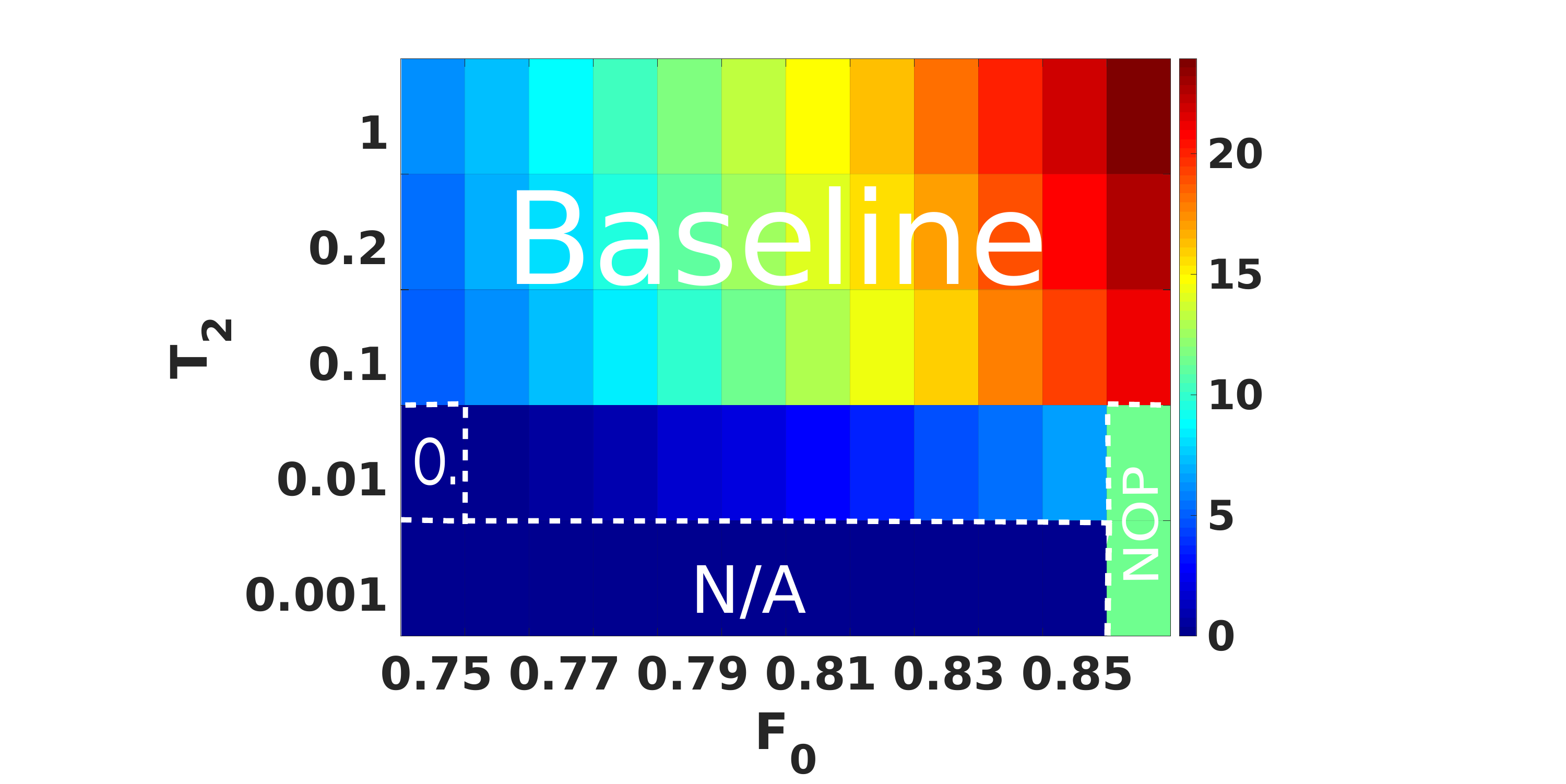}
  \label{fig:bb84_stefan_1}}\qquad
  \subfigure[Ground-based setup with 1 MHz EPR source rate ($\mu$) and $d=20$ km. ]{\includegraphics[scale=\x]{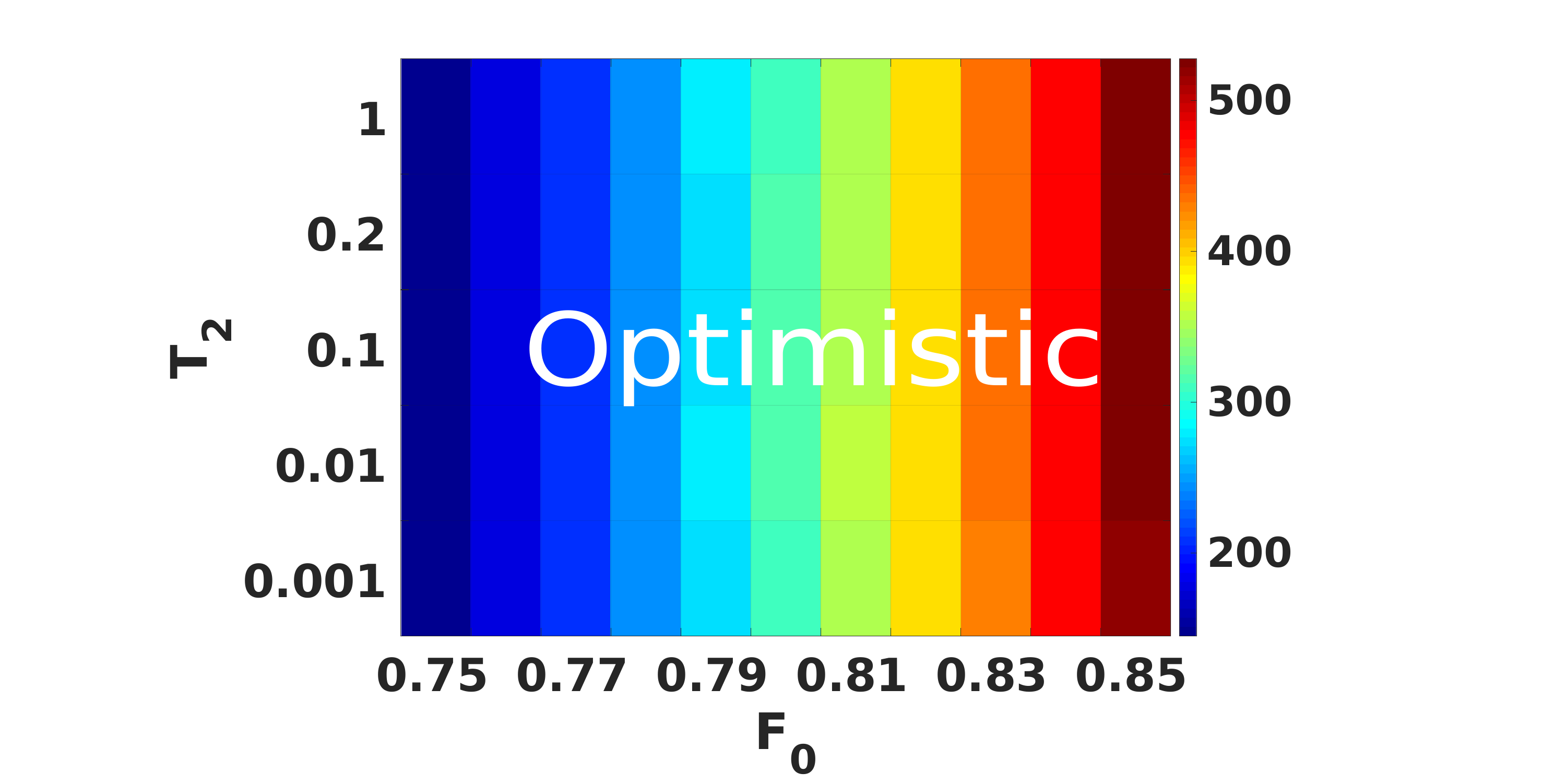}
    \label{fig:bb84_stefan_2}}\qquad
  \subfigure[Satellite-based setup with 1 KHz EPR source rate ($\mu$) and $d=500$ km. ]{\includegraphics[scale=\x]{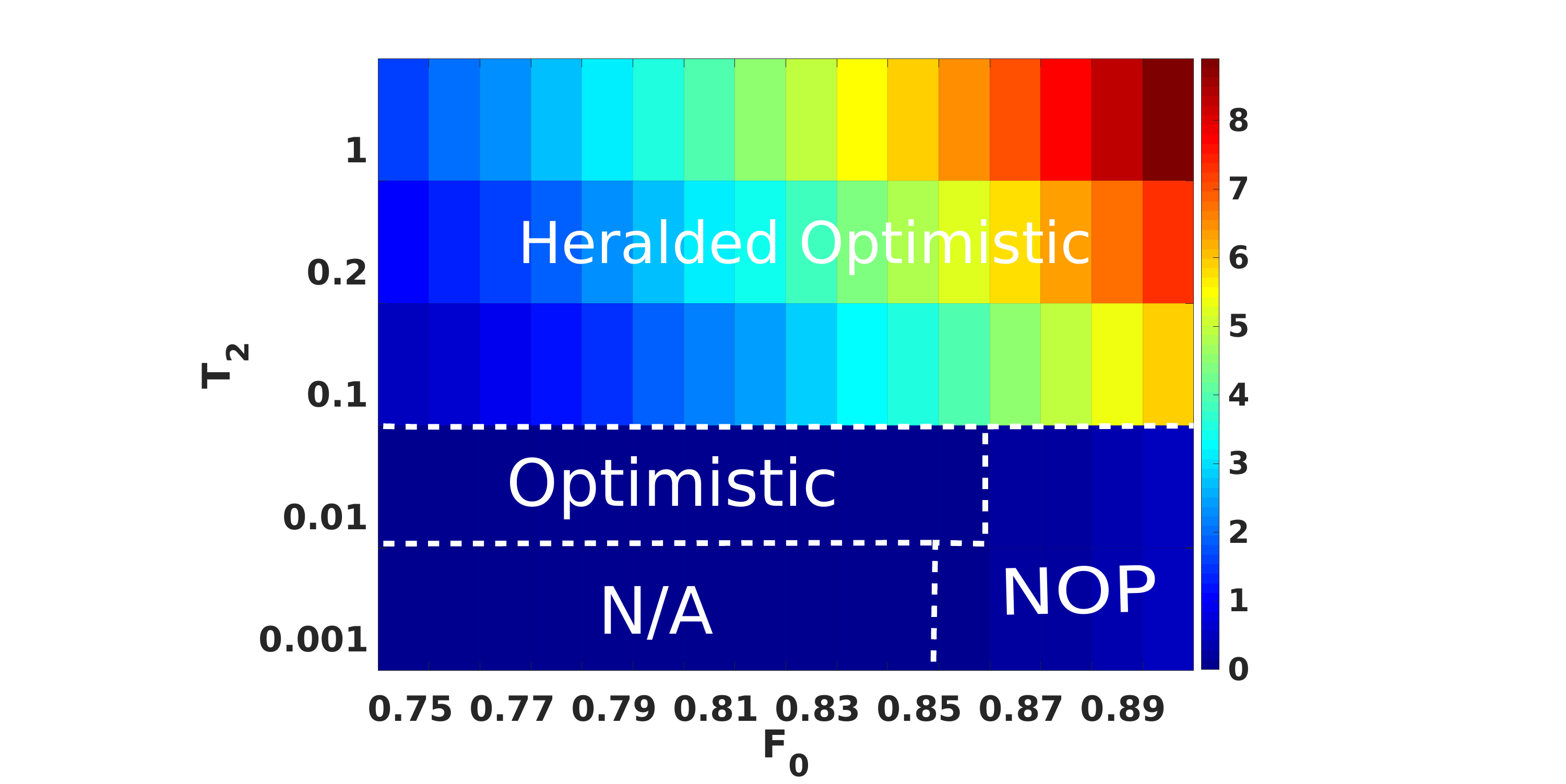}
    \label{fig:bb84_stefan_3}}\quad
  \subfigure[Satellite-based setup with 
 1 MHz EPR source rate ($\mu$) and $d=500$ km. ]{\hspace{4mm}\includegraphics[scale=\x]{    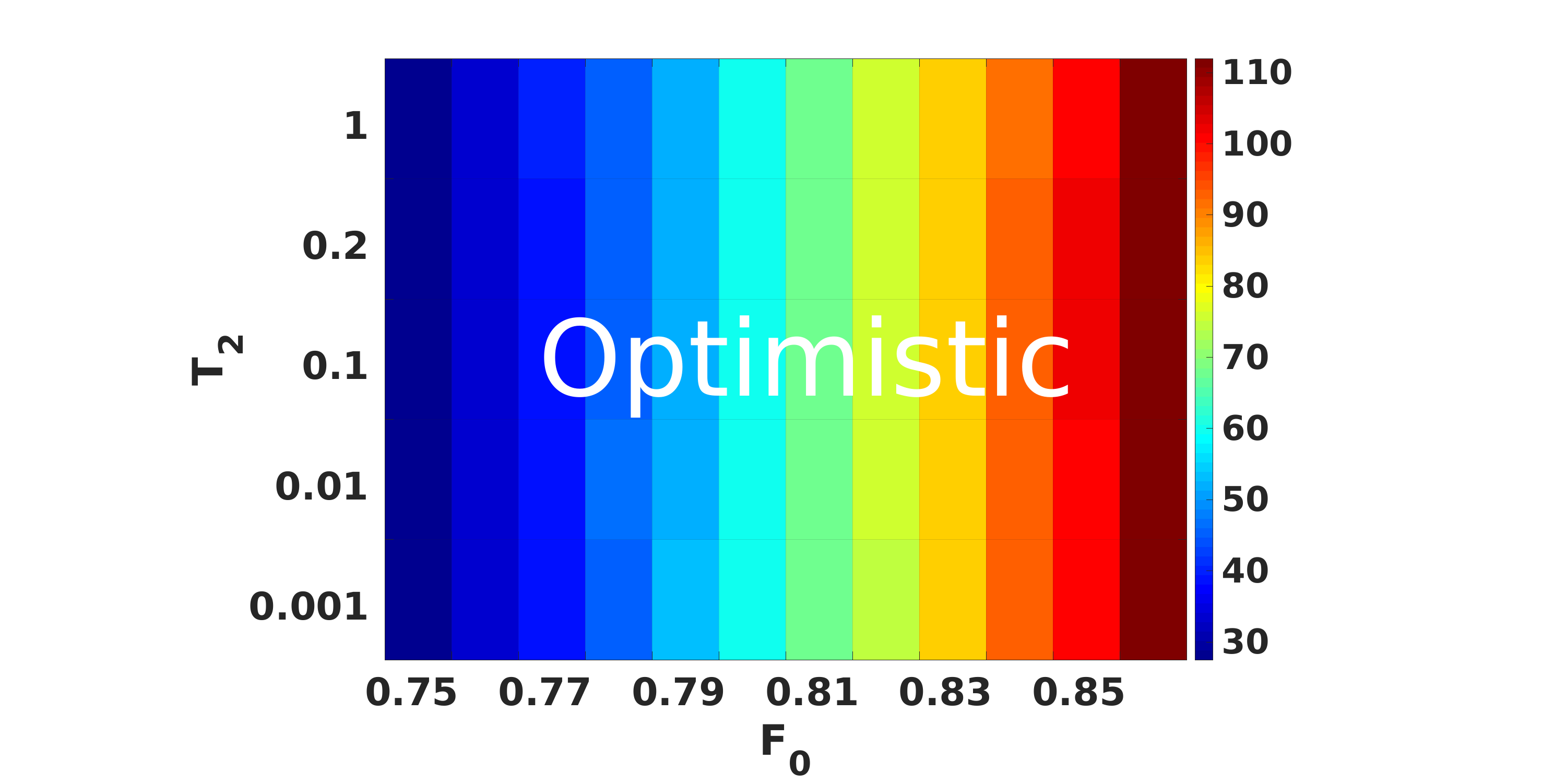}
    \label{fig:bb84_stefan_4}}
\caption{Heatmaps for BB84 for an optimized purification circuit as a function of $T_2$ and $F_0$ for various rates and EPR pair generation setups. 
 We demarcate different regions with dashed lines and label them to indicate the protocol with the highest QKD SKR for each region. `NOP' indicates generating secret keys without any purification and `O.' stands for optimistic. We indicate the regions where SKR is zero with `N/A' label.}

\vspace{-10pt}
\label{fig:bb84_stefan}
\end{figure*}
\subsection{Pumping Scheme}
In this section, we evaluate the effect of the number of purification steps on average fidelity and rate. Then we study the effect of distance ($d$) on average fidelity. Next, we study the effect of coherence time ($T_2$) and EPR source rate ($\mu$) on average fidelity. We compare OPT with HOPT, BASE, and NOP in a ground-based setup. For all cases, we do at most five steps of purification in the pumping scheme as going further does not improve fidelity significantly. We limit the number of memories to two for each node, the required number of memories for each step of the pumping scheme. We set $F_0 = 0.9$, $\mu = 1$ GHz, $T_2 = 0.001$ s, and $d = 20$ km; and plot the results for fidelity and rate in Figure~\ref{fig:fid_pumping} and Figure~\ref{fig:rate_pumping} respectively. We observe in Figuer~\ref{fig:fid_pumping} that OPT outperforms all other protocols, while BASE and HOPT yield fidelities lower than NOP. We also observe in Figure~\ref{fig:rate_pumping} that OPT yields a lower rate compared to other protocols, however, the higher rates of HOPT and BASE  do not compensate for their lower fidelity.
Next, we study the effect of $d$ on average output fidelity. We plot the highest fidelity achieved over five steps of purification in Figure~\ref{fig:fid_vs_dist}, note that for OPT, step five always has the highest fidelity, while for BASE or HOPT this may not be the case (see Figure~\ref{fig:fid_pumping}). NOP is also included in the plot.  We find that OPT outperforms other protocols, and by increasing $d$, the difference between OPT and other purification protocols increases. Furthermore, an increase in $d$ leads to a longer waiting time, which results in a decrease in the fidelity of both BASE and HOPT, causing them to fall below the fidelity of NOP. Moreover, for larger $d$, the difference between HOPT and BASE decreases as they perform fewer purification steps and when they decrease to one step their performance becomes the same. 

We next analyze the effect of $T_2$ on the average fidelity of all protocols for $F_0 = 0.9$, $d = 5, 20$ km, and $\mu = 1$ GHz.
To investigate the impact of $T_2$, we compare average output fidelity across all protocols for different values of this parameter.  We plot the highest average fidelity that is achieved over the number of purification steps as a function of $T_2$ for all protocols in Figure~\ref{fig:varying_t2_dejmps}.
We observe that by increasing $T_2$, all purification protocols converge to the same output fidelity. 
The distance between nodes plays a role in the convergence behavior: for 5 km, the difference between different protocol fidelities is negligible for $T_2$ larger than $0.01$ s, and for 20 km, when $T_2$ is larger than $0.1$ s. 

EPR source rate, $\mu$, also affects output fidelity, and consequently the selection of a purification protocol. We study the effect of $\mu$ on average fidelity for $F_0 = 0.9$, $d = 5,20$ km, and $T_2 = 1$ ms and plot it in Figure~\ref{fig:varying_rate_dejmps}. As $\mu$ decreases, each qubit spends more time in memory and therefore is subjected to decoherence for a longer period. By increasing $\mu$, output fidelity increases; however, when the rate surpasses 1 MHz, output fidelity improvement is negligible for all protocols.

\subsection{Satellite-based EPR Generation Setup}
In this section, we evaluate the performance of all purification protocols, implementing a pumping scheme, in a satellite-based setup. Similar to the ground-based setup, the maximum number of purification steps is five, as beyond five steps the fidelity does not improve significantly. We restrict each node to have a maximum of two memories, the minimum number of memories for each step of the pumping scheme. Since the distances between two end nodes are significantly larger compared to the ground-based setup, memory decoherence becomes more severe. Consequently, having a memory with a coherence time of $1$ ms significantly reduces overall fidelity due to the large waiting time induced by classical communication; we therefore set the coherence time to $10$ ms for fidelity and rate evaluation. We plot the average fidelity as a function of number of purification steps in Figure~\ref{fig:sat_fidelity} for $F_0 = 0.9$, $\mu = 1$ GHz, $d = 500$ km, and $T_2 = 10$ ms.
We observe that OPT provides the highest average fidelity, while HOPT and BASE yield lower fidelities compared to NOP. For the same parameters, we plot the average entanglement rate as a function of number of purification steps in Figure~\ref{fig:sat_rate}. We observe that the average entanglement rate under OPT is lower than other protocols; however, the average entanglement rates of HOPT and BASE are lower than NOP. This brings up the rate-fidelity trade-off problem, which we study in Section~\ref{section:eval_qkd}. \\

Next, we explore the effect of inter node distance $d$ on average fidelity in Figure~\ref{fig:sat_fidelity_distance}. We observe that OPT exhibits superior performance compared to the other protocols, and as $d$ increases, the performance gap widens. In addition, an increase in $d$ leads to a longer EPR pairs storage time in noisy memories, which results in a decrease in the performance of BASE and HOPT, causing them to fall below that of NOP.

We then analyze the effect of $T_2$ and $\mu$ on the average fidelities of all protocols. To study the effect of $T_2$, we fix $\mu$ to 1 GHz and $F_0$ to 0.9 for distances of 200 and 500 km. We plot average fidelity as a function of $T_2$ in Figure~\ref{fig:sat_varying_t2}. We observe that OPT yields the highest fidelity, but that, as $T_2$ increases, all purification protocols converge to the same fidelity value and once $T_2 = 0.5\,s$ the difference is negligible.
For evaluating the effect $\mu$ on the average fidelity, we set $F_0 = 0.9$ and $T_2 = 10$ ms. We plot average fidelity as a function of $\mu$ in Figure~\ref{fig:sat_varying_rate}, where we observe that increasing $\mu$ improves average fidelity and that after $ \mu = 1$ MHz this improvement is negligible.\\
Our analysis shows a significant decrease in final average fidelities in the satellite scenario due to large classical communication latencies. For example in Figure~\ref{fig:sat_fidelity}, we observe that OPT only increases fidelity to $\sim 0.75$, which is not suitable for many applications. Some applications such as QKD~\cite{paper:qkd,paper:qkd_e91,paper:qkd_secret_key_fraction}, allow end nodes to measure their qubits as soon as they receive or purify them and continue to receive EPR pairs and/or purify them; and at the meantime, exchange the heralding and purification results to see whether purification was successful or not to filter out failed EPR pairs. We calculate fidelity and the rate with the modification in which end nodes measure their qubits before the final confirmation.
We show fidelity and rate as a function of number of purification steps for the same hardware parameters set previously in Figure~\ref{fig:sat_fidelity_w/o} and Figure~\ref{fig:sat_rate_w/o}, respectively. We observe that average fidelity increases due to the decrease in waiting time. This reduction in wait time also improves the overall rates of all protocols and amplifies OPT's rate to surpass those of BASE and HOPT.
\subsection{Optimized Purification Circuit}
We now evaluate the benefit of optimism in the context of a circuit generated by a genetic algorithm introduced in~\cite{paper:purification_last_stefan}. To do so, we remove amplitude-damping noise in quantum memories since the genetic algorithm of \cite{paper:purification_last_stefan} does not support this noise model. However, as discussed previously, we do not expect this to have a significant impact on results since qubits are stored in memories for relatively short periods of time, on the order of milliseconds. For evaluation, we use the genetic algorithm to produce an optimized circuit similar to the original L17 circuit of \cite{paper:purification_last_stefan} that has the same performance in terms of fidelity improvement and average number of consumed EPR pairs. The circuit has 17 operations and requires nine EPR pairs and three quantum memories. We selected L17 as the basis of our design since it outperforms the STRINGENT protocol~\cite{paper:optimized_stringent}.
To evaluate fidelity and rate, we set $\mu = 1$ GHz and $T_2 = 0.001$ s, the same as the ground-based pumping scheme evaluation.
Average fidelity and average rate as a function of $d$ can be found in Figure~\ref{fig:optimized_fidelity} and Figure~\ref{fig:optimized_rate}, respectively. 
We find that the OPT outperforms other protocols in terms of fidelity. Further, BASE and HOPT yield lower fidelities than NOP for longer distances. OPT achieves higher rates than the other protocols for $d < 4.3$ km.

Next, we evaluate the effect of $T_2$ and $\mu$ on average fidelity for a 5 km link.  We plot average fidelity as a function of $T_2$ in Figure~\ref{fig:varying_t2_optimized} for $F_0 = 0.9$ and $\mu = 1$ GHz.  As $T_2$ increases, the fidelity difference between OPT and other protocols decreases, becoming negligible for $T_2 > 0.1$ s. We also study the effect of $\mu$ on the average fidelity in Figure~\ref{fig:varying_rate_optimized}. We observe that by increasing $\mu$,  average fidelity improves, and when $\mu$ surpasses 1 MHz, this improvement is negligible.
\vspace{-0.1cm}
\subsection{Secret Key Rate Evaluation}
\label{section:eval_qkd}
In this section, we study the rate-fidelity trade-off for all protocols by evaluating their performance in the context of BB84's SKR~\cite{paper:qkd_secret_key_fraction}. In previous descriptions (see Figures~\ref{fig:baseline} and \ref{fig:optimistic}), all protocols wait for the final confirmation and purification results, and users do not receive new EPR pairs while waiting. In the case of QKD, we make the modification that end nodes measure the EPR pair of a purification procedure prior to the final confirmation so that the measurement output can be sent along with purification and heralding results; this way, their memories are free and able to receive new EPR pairs, allowing the generation of the next secret key bit to proceed. This modification for QKD yields the greatest benefit where distances between end nodes are large, such as a satellite setting. 
Similar to previous sections, we study the effect $T_2$, $d$, $F_0$, and $\mu$ on the secret key rate in ground-based and satellite-based scenarios. We evaluate QKD performance for the pumping scheme and the optimized purification circuit of~\cite{paper:purification_last_stefan}. In our simulations, we set $d$ to $20$ and $500$ km for ground- and satellite-based settings, respectively. We set $\mu $ equal to 1 KHz and 1 MHz for both scenarios (for the evaluation of 10 KHz and 100 KHz of pumping scheme see Appendix~\ref{FirstAppendix}). In all cases, we consider a range of values for $T_2$ and initial fidelity $F_0$. We display the maximum SKR across all protocols for each combination of $F_0$ (x-axis) and $T_2$ (y-axis). 
We partition different regions of the heatmap with dashed lines to indicate which protocol achieved the maximum SKR for each $(F_0, T_2)$ pair.

For entanglement pumping, presented in Figure~\ref{fig:pumping_bb84}, our study indicates that at  $T_2 = 0.01$, $F_0$ in the range of 0.79 to 0.85, and $\mu = 1$ KHz, OPT outperforms other variants. Increasing $T_2$ and $F_0$ improves the performance of BASE and HOPT (see Figure~\ref{fig:pumping_bb84_1} for ground-based setup and Figure~\ref{fig:pumping_bb84_3} for satellite-based setup). By increasing $\mu$ to 1 MHz, the OPT approach outperforms other approaches for all $(F_0, T_2)$ (see Figure~\ref{fig:pumping_bb84_2} for ground-based setup and Figure~\ref{fig:pumping_bb84_4} for satellite-based setup).

For the purification circuit of~\cite{paper:purification_last_stefan}, we modified the fitness function of the genetic algorithm to generate a new circuit optimized to the SKR of BB84 (the original algorithm's fitness function aims at maximizing the output fidelity). Moreover, we consider storage noise while generating the circuit. We generate a circuit that uses three memories with $T_2$ of 0.01, requires five EPR pairs of initial fidelity 0.75, and has at its disposal a 1 KHz EPR source. Figure~\ref{fig:bb84_stefan} presents the performance of the circuit for all three purification protocols.
Similar to the pumping scheme, we observe that OPT outperforms other protocols when $F_0$ and $T_2$ are low(see Figures~\ref{fig:bb84_stefan_1} and \ref{fig:bb84_stefan_3} for ground-based and satellite-based setups respectively). It is worth mentioning that our generated circuit outperforms the pumping protocol, in that it is capable of achieving a positive SKR using OPT, in cases where $F_0$ and $T_2$ are so low that the pumping protocol can not yield a key. For example, Figure~\ref{fig:bb84_stefan_1}, for $T_2 = 0.01$ and $ F_0  = 0.75$, shows that the optimized circuit utilizing our optimistic protocol can generate a secret key, while pumping cannot generate any secret key (see SKR for $T_2 = 0.01$ and $ F_0 \leq  0.78$ in Figure~\ref{fig:pumping_bb84_1}). Similar to the pumping scheme, by increasing the EPR source rate to 1 MHz, the optimistic protocol outperforms other protocols for all $(F_0, T_2)$ (see Figures~\ref{fig:bb84_stefan_2} and \ref{fig:bb84_stefan_4} for ground- and satellite-based setups respectively).
\section{Conclusion and Future Directions}
\label{section:conclusion}
In this work, we proposed optimism in purification circuits. Our study showed that being optimistic about heralding signals and purification results can be advantageous to fidelity and, in some hardware parameter regimes, to overall purified EPR rate in classic purification schemes (\textit{e.g.}, entanglement pumping) and optimized purification circuits of\cite{paper:purification_last_stefan}, compared to baseline (\textit{i.e.}, herald all EPR pairs, check every purification result) and heralded-optimistic (\textit{i.e.}, herald EPR pairs, exchange purification results only while heralding) approaches. We study the effects of memory and gate noise; EPR source rate, and node distance on the performance of our proposed optimistic protocol and compare it to the aforementioned protocols. As part of a future direction, we aim to evaluate our proposed scheme on real hardware such as NV centers in diamond~\cite{paper:nv_center, paper:linklayer, paper:netsquid}. Moreover, we aim to test our approach on a quantum repeater chain and analyze the effect of different parameters on the output fidelity and overall end-to-end EPR rate. 
The optimistic approach can also be applied to GHZ state~\cite{paper:ghz} distribution schemes. In~\cite{paper:ghz_first} authors proposed a procedure to distribute a quadripartite GHZ state between four end nodes. This involves generating four Bell pairs and applying purification, then applying a procedure called fusion (for an optimized version of GHZ distribution and fusion see ~\cite{paper:ghz_optimized}) to generate the desired quadripartite GHZ. We expect that for such a task, the optimistic approach would benefit both fidelity and rate.

\section*{Acknowledgment}\vspace{-0.05in}
This research was supported in part by the NSF grant CNS-1955744, NSF-ERC Center for Quantum Networks grant EEC-1941583, and MURI ARO Grant W911NF2110325.

\bibliographystyle{IEEEtran}
\bibliography{bare_conf}

\appendices
\begin{figure}[t]
\newcommand\x{0.17}
  \centering
  \hspace{-4mm}\subfigure[Ground-based setup with 10 KHz EPR source rate ($\mu$) and $d=20$ km.]{\includegraphics[scale = \x]{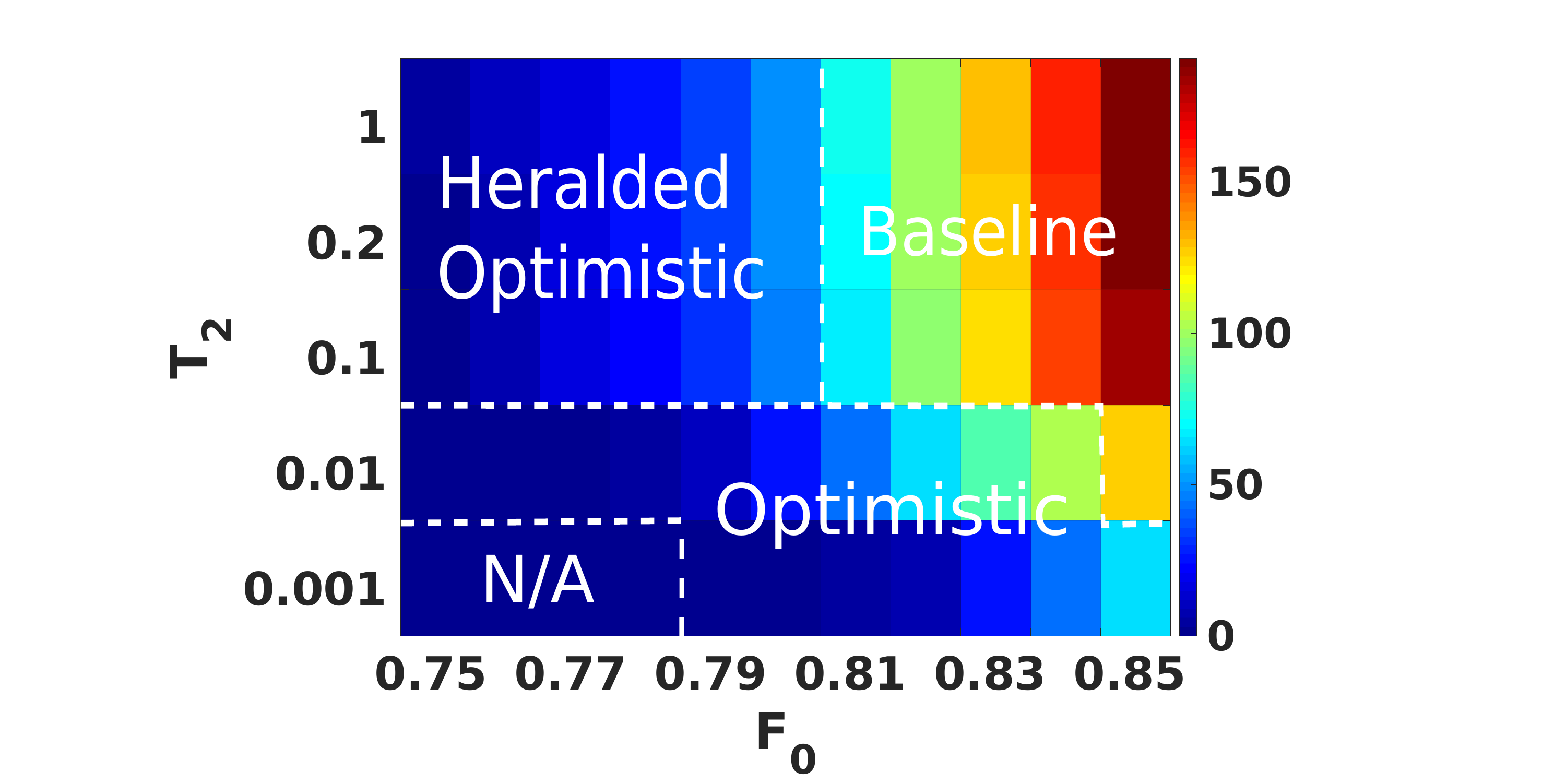}
  \label{fig:pumping_bb84_1_appendix}}\qquad
  \subfigure[Ground-based setup with 100 KHz EPR source rate ($\mu$) and $d=20$ km.]{\hspace{-4.5mm}\includegraphics[scale = \x]{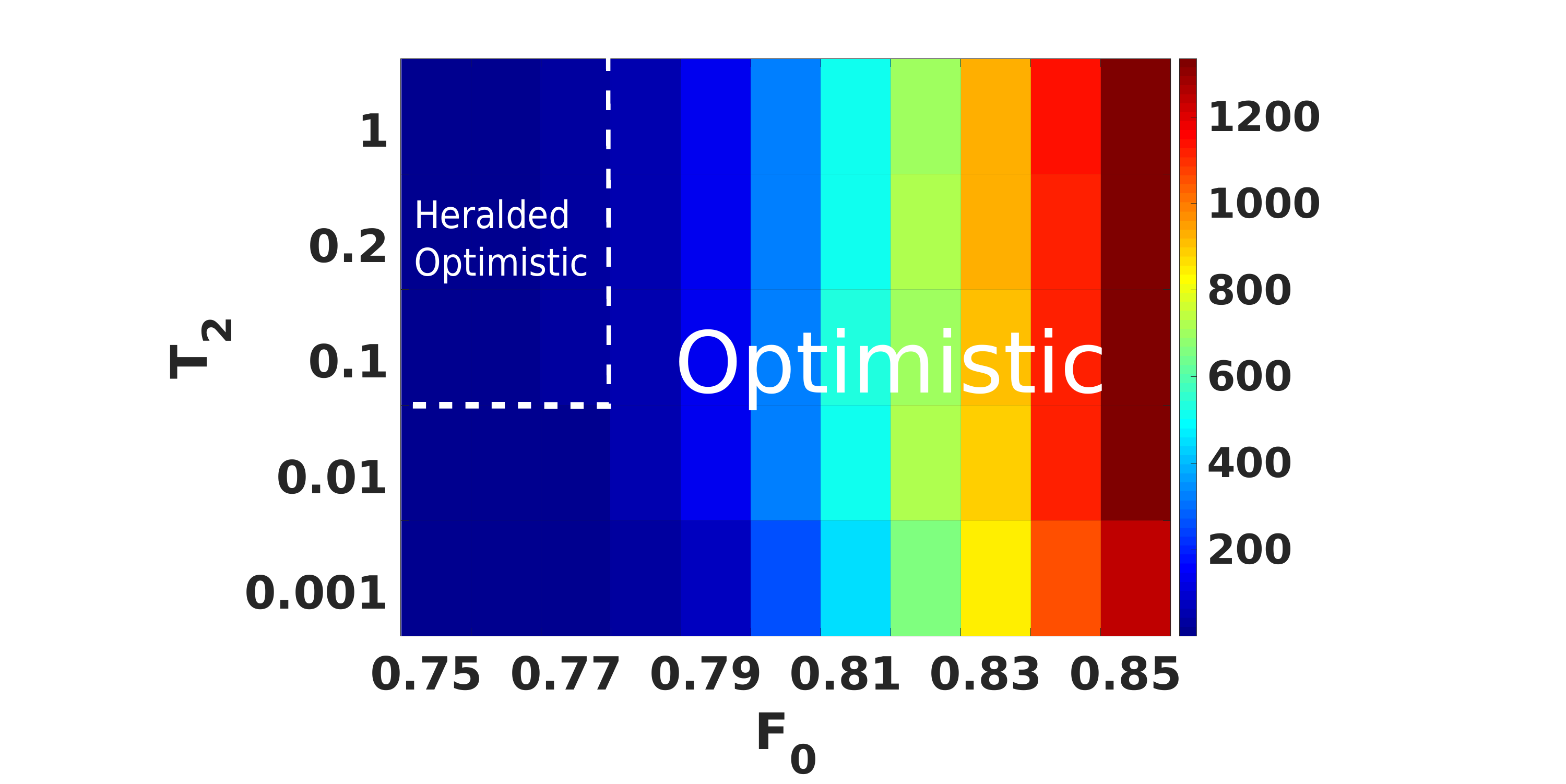}}
    \label{fig:pumping_bb84_2_appendix}\qquad
 \hspace{-12mm} \subfigure[Satellite-based setup with 10 KHz EPR source rate ($\mu$) and $d=500$ km.]{\hspace{1mm}\includegraphics[scale=\x]{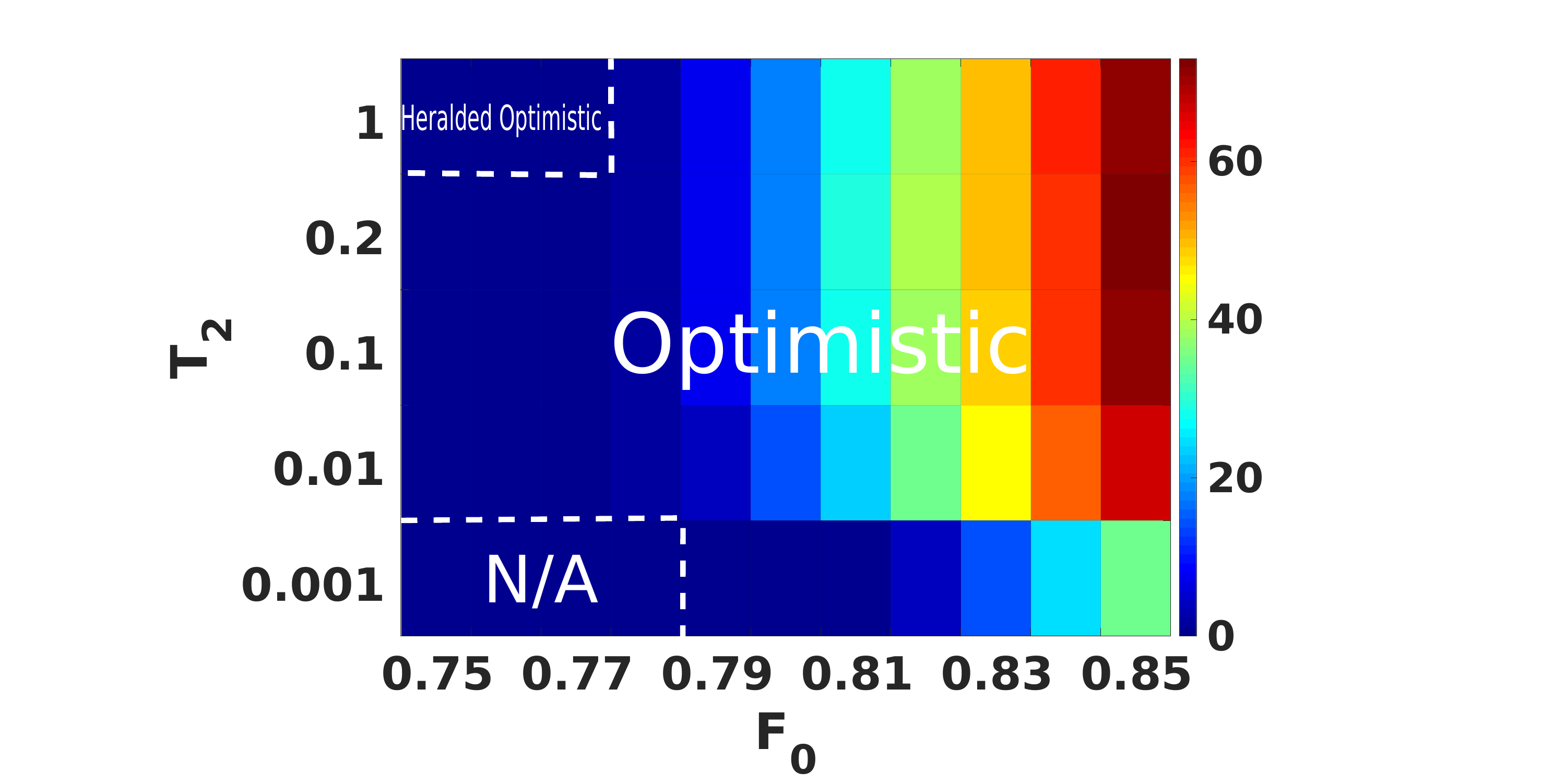}
\hspace{-5mm}    \label{fig:pumping_bb84_3_appendix}}\qquad
  \subfigure[Satellite-based setup with 100 KHz EPR source rate ($\mu$) and $d=500$ km.]{\includegraphics[scale=\x]{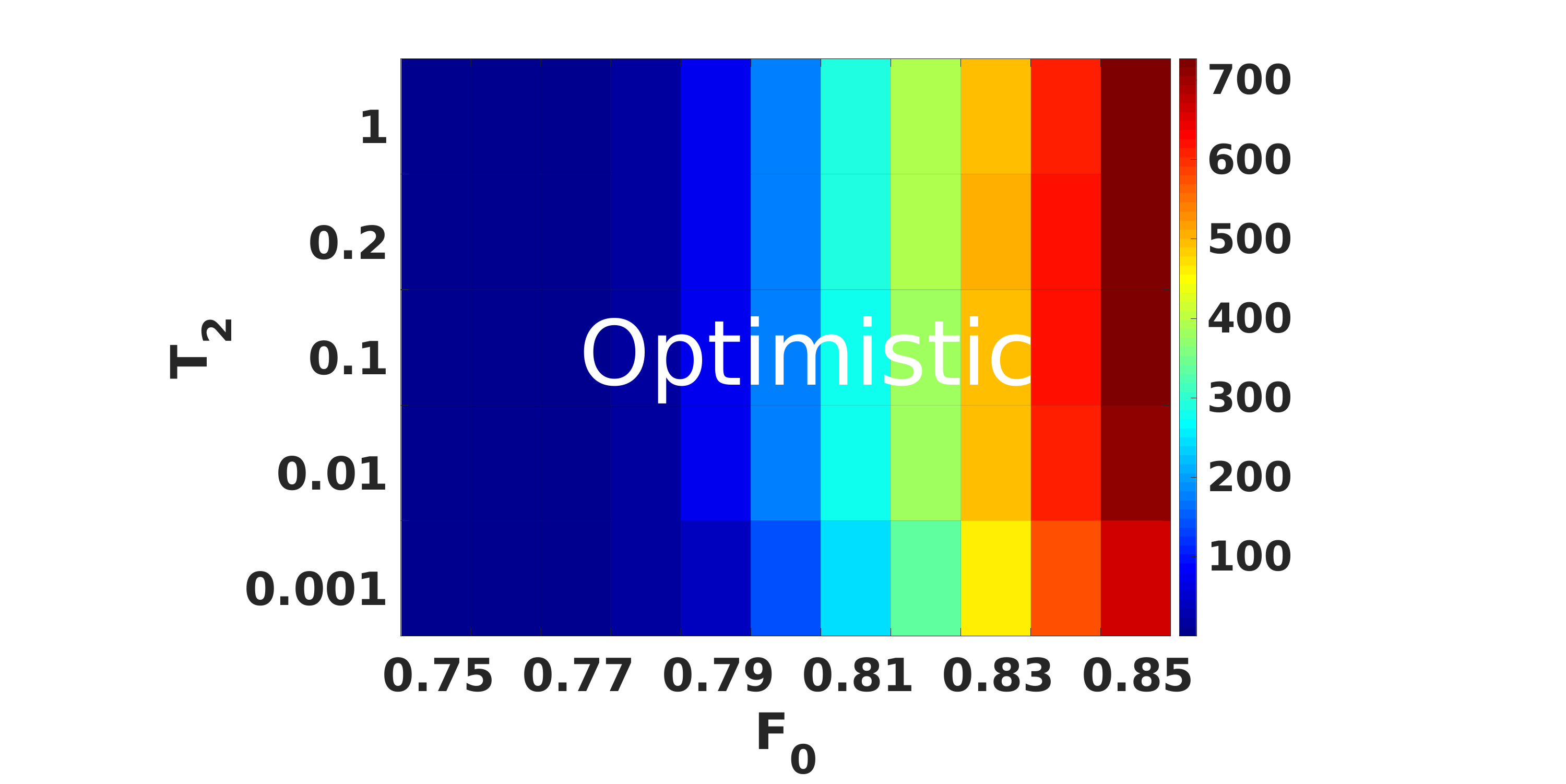}
    \label{fig:pumping_bb84_4_appendix}}
\caption{Heatmaps for BB84 SKR using the pumping scheme, as a function of $T_2$ and $F_0$ for various rates and EPR pair generation setups. We demarcate different regions with dashed lines and label each region to show the best protocol for QKD in that region. The `N/A' label indicates values of ($F_0$, $T_2$) where no positive SKR can be achieved.}
\vspace{-8pt}
\label{fig:pumping_bb84_appendix}
\end{figure}
\section{}
\label{FirstAppendix}

In this appendix, we plot the SKR for purification protocols with an EPR source rate ($\mu$) of 10 and 100 KHz using a pumping scheme for ground-based and satellite-based setups. The results are shown in Figure~\ref{fig:pumping_bb84_appendix} in the form of a heatmap plot (see Section~\ref{section:eval_qkd}). We observe that by increasing $\mu$, OPT outperforms other protocols in more regions. The reason behind this improvement is that with a higher $\mu$, OPT can receive more EPR pairs. In contrast, HOPT and BASE cannot, because they have to wait for more confirmation throughout the process (HOPT for heralding, and BASE for heralding and purification confirmation), preventing them from receiving and performing operations on EPR pairs.

\end{document}